\documentclass[11pt,a4paper]{iopart}
\usepackage{iopams}
\usepackage{setstack,cite}
\usepackage{amssymb,graphicx}
\usepackage{braket}

\newcommand{\bea}{\begin{eqnarray}}
\newcommand{\eea}{\end{eqnarray}}
\newcommand{\bes}{\numparts}
\newcommand{\ees}{\endnumparts}

\newcommand{\sn}{{\rm sn}}
\newcommand{\dn}{{\rm dn}}
\newcommand{\cn}{{\rm cn}}
\newcommand{\sech}{{\rm sech}}

\begin{document}
\bibliographystyle{iopart-num}
\title{Spatially modulated two- and three-component Rabi-coupled Gross-Pitaevskii systems}
\author{T Kanna$^1$\footnote{kanna\_phy@bhc.edu.in (corresponding author)}, A Annamalar Sheela$^1$ and R Babu Mareeswaran$^1$}
\address{$^1$ Nonlinear Waves Research Lab,  PG and Research Department of Physics,\\ Bishop Heber College, Tiruchirapalli 620 017, Tamil Nadu, India.}

\begin{abstract}
	Vector rogue wave (RW) formation and their dynamics in Rabi coupled two- and three-species Bose-Einstein condensates with spatially varying dispersion
	and nonlinearity are studied. For this purpose, we obtain the RW solution of the two- and three-component inhomogeneous Gross-Pitaevskii (GP)
	systems with Rabi coupling by introducing suitable rotational and similarity transformations. Then, we investigate the effect of inhomogeneity (spatially
	varying dispersion, trapping potential and nonlinearity) on vector RWs for two different forms of potential strengths, namely periodic (optical lattice) with specific reference to hyperbolic type potentials and parabolic cylinder potentials. First, we show an interesting oscillating boomeronic behaviour of dark-bright solitons due to Rabi coupling in two component-condensate with constant nonlinearities. Then in the presence of inhomogeneity but in the absence of Rabi coupling we demonstrate creation of new daughter RWs co-existing with dark (bright) soliton part in first (second) component of the two-component GP system. Further, the hyperbolic modulation (sech type) of parameter along with Rabi effect leads to the formation of dromion (two-dimensional localized structure) trains even in the (1+1) dimensional two-component GP system, which is a striking feature of Rabi coupling with spatial modulation. Next, our study on three-component condensate, reveals the fact that the three RWs can be converted into broad based zero background RW appearing on top of a bright soliton by introducing spatial modulation only. Further, by including Rabi coupling we observe beating behaviour of solitons with internal oscillations mostly at the wings. Also, we show that by employing parabolic cylinder modulation with model parameter $n$, one can produce $(n+1)$ RWs. Our study reveals the fact that the spatially varying environment leads to the possibility of realization of two dimensional  nonlinear structure in (1+1) dimensional setting.
\end{abstract}
\section{Introduction}
The past two decades have witnessed a growing interest in rogue wave (RW) research \cite{kharif2009,osborne2009,guo2017,akhmediev2013}. These RWs are indeed giant nonlinear waves that appear from nowhere and disappear without a trace \cite{ankiewicz2009}. It is generally recognized that modulation instability is one of the prime mechanisms leading to the formation of RW excitation \cite{kharif2009,solli2007}. The versatile nonlinear Schr\"{o}dinger (NLS) equation is a prototype nonlinear equation that features such RWs. Study of RWs has attracted significant attention not only in the more standard ocean-surface-dynamical problem \cite{kharif2009,osborne2009},
but also in other physical contexts. Indeed, there exists a vast amount of theoretical and experimental works on RWs in various fields ranging from optics
\cite{akhmediev2013,solli2007,xu2011,lecaplain2012,kibler2012}, Bose-Einstein condensates (BECs) \cite{kharif2009,chabchoub2012},
to plasmas \cite{moslem2011}, and atmospheric dynamics \cite{stenflo2010} (see also the recent short review \cite{onorato2013}). Recently,
high-order RWs \cite{erkintalo2011} were excited successfully in a water wave tank. This suggests that a high-order analytic RW solution is
of physical relevance and can be realized experimentally \cite{erkintalo2011,chabchoub2013}.

These comprehensive theoretical and experimental studies on single component RWs, stimulated an increasing interest in investigating multi-component RWs. Rogue waves of the two-component Manakov system \cite{manakov1974} have been studied extensively in the focusing regime \cite{baronio2012,ling2014} and in the defocusing regime \cite{bludov2010,degasperis2013,zhao2012}. The higher order RW solutions of the Manakov system have also been constructed by employing the Darboux transformation (DT) method  \cite{ling2013}. Furthermore, RWs for the three-component NLS system in the focusing regime
features interesting four-petal pattern RW \cite{zhao2013,liu2014}. Very recently the single, doublet, triplet and quadruple RWs of a fairly general M-component NLS system has been obtained by using Kadomstev-Petviashvili hierarchy reduction method \cite{rao2019}. Such wave profiles are otherwise not possible in the single and two component NLS systems. These studies indicate that there exists very abundant pattern dynamics for RWs in the multi-component nonlinear systems,
which is quite distinct from those of single component (scalar) nonlinear dynamical systems.

Bose-Einstein condensates act as a fertile arena for theoretical and experimental realization of multi-component nonlinear systems.
These multi-component BECs consist of atoms in different internal states or even of different species. There are several experiments related to
such multi-component condensates \cite{myatt1997,hall1998,stamper1998}.  The interesting experiment \cite{kartashov2011a} on $^{87}$Rb shows
that, the irradiation of the condensate with an electric field induces a linear coupling (proportional to the Rabi frequency) in the overlap region,
which causes {\it Rabi oscillations} between the two components. This Rabi coupling transfers an arbitrary fraction of atoms from one component to
another component, e.g., from the $\ket{1, −1}$ state to the $\ket{2, 1}$ state. 
 In Ref. \cite{inouye1998}, it was shown that in analogy with systems
arising in the field of nonlinear fiber optics (such as a twisted fiber with two linear polarizations), exact Rabi oscillations between two
condensate can be analytically found when inter-species coupling being equal to unity. Further, Ref. \cite{fatemi2000,deconick2004}, generalized this
concept to multi-component BECs with time varying Rabi coupling and shows that Rabi switch is very robust with high efficiency when transferring
a condensate wave function between the components.

The techniques for managing nonlinearity (spatial or temporal dependence of nonlinearity) to produce nonlinear waves in BECs have also gained
considerable attention \cite{malomed2006}. Specifically, the Gross-Pitaevskii (GP) equation describing the dynamics of BEC in the mean field approximation has been examined
in the presence of temporally \cite{saito2003} or spatially \cite{kartashov2011a,zhong2014,zhong2015}  varying nonlinearity coefficients.
Such spatial or temporal variations of nonlinearities can be realized experimentally very well in the atomic condensate by employing the Feshbach resonance mechanism, where the $s$-wave scattering length is tuned in space or time with the aid of external magnetic \cite{inouye1998} or optical \cite{fatemi2000} fields (for a review see Ref. \cite{chin2010}). In particular, much attention has been paid to spatially modulated nonlinearities in BECs which lead to collisionally inhomogeneous condensates \cite{maria2005,theocharis2005,middelkamp2009,ran2011,khamehchi2017}. Recently, the scattering of such matter wave solitons at an interface has been studied in \cite{tsitoura2015}. Then the dark solitons in such spatially varying environment with linear, nonlinear potential steps and double potential barrier have been studied in Refs. \cite{tsitoura2016} and \cite{tsitoura2017} respectively. Very recently, single as well as multi-peak two dimensional (2D) soliton and vortex solitons are studied in parity-time ($\cal PT$) symmetric atomic and condensates with spatially modulated nonlinearities \cite{eitam2019}.

On the other hand, the experimental demonstration of controlling dispersion of matter wave packet by engineering the effective mass \cite{prl2003} and the observation of variation of the sign of the effective mass of BEC from positive to negative when the acceleration is no longer proportional to the force \cite{khamehchi2017,colas2018} suggest the possibility of varying mass also along the spatial direction in the description of BEC. This varying mass will ultimately lead to the coefficient of the kinetic energy term (spatial dispersion coefficient) in the GP system responsible for the dispersion of the matter wave to be a function of spatial co-ordinates. From another experimental view point, spatial variation of mass can also be achieved by imposing a nonuniform optical lattice created by laser beams over a condensate in a cigar-shaped trapping potential \cite{brazhnyi2004}. In addition to atomic condensates, the NLS type equations with trapping potential, varying dispersion and nonlinearity (referred as variable coefficient GP system) have been studied extensively in the context of nonlinear optics, especially in dispersion managed systems \cite{turitsyn2012} and the notion of nonlinearity management was also theoretically developed \cite{towers2002} that has been well established experimentally later on \cite{centurion2006}.

These fascinating concepts of varying dispersion and nonlinearity management can be profitably employed to stabilize RWs by making their corresponding coefficients to be functions of the propagation distance or transverse coordinate or both. One of the mathematical approaches to deal with such problems
is to employ a similarity transformation which is indeed a nonlinear transformation
that transforms the original variable coefficient nonlinear evolution equation into an integrable system that admits soliton, breather, rogue wave solutions etc. \cite{serkin2007,belmonte2007,rajendran2011,yan2010,he2011,xu2012,he2013,kanna2013}. This method works very well in higher dimensions too \cite{yan2009,li2018}. By this way, the exact RW solutions have been obtained in several (1+1) dimensional scalar and vector GP systems by choosing appropriate forms of trapping potential and nonlinearity coefficients \cite{bludov2010,zhong2015,he2014,babu2014,kanna2013,kanna2017} but with constant coefficient. Especially, dynamics of bright-bright, bright-dark solitons and vector boomeronic solitons in non-autonomous Rabi coupled GP system are discussed in Ref.\cite{babu2014,kanna2013} in a detailed manner. Later, the tuning mechanism of dispersion along with nonlinearity is employed to study RWs in the Manakov system with spatially varying coefficients for a parabolic cylinder potential \cite{zhong2015}. Subsequently, the binary condensates in a three dimensional setting with spatially varying nonlinearity is considered in Ref. \cite{xu2018} and vector solitary waves are obtained along with a stability analysis. As a next step, it is  quite natural to pose the question how the spatial variation of the nonlinearity and dispersion influences the RW phenomenon in Rabi coupled condensates. This study will also be relevant to pico-second pulse propagation in tapered twisted birefringent fibers as the governing equation for both these settings are identical.

Being motivated by the above reasons, in this work we study the multi-component BECs, particularly two- and three-component BECs, with spatially dependent dispersion coefficient, scattering length and trapping potential by including Rabi coupling.

The rest of this paper is organized as follows. In section 2, we introduce two successive transformations and reduce the two-component inhomogeneous (spatially modulated) GP system (\ref{model}) to the canonical Manakov system and discuss the general RW solutions by considering two different types of trapping potentials namely optical lattice and parabolic cylinder potentials. Then, in section 3, we consider the three-component inhomogeneous GP equation with Rabi coupling and explore interesting dynamical features of RWs. Finally, we conclude our study in section 4.

\section{Two-component GP equation with Rabi coupling}

\subsection{Description of the system}
To start with, we consider a spatially modulated two-component GP (2-CGP) equation with Rabi coupling following Refs. \cite{zhong2014,zhong2015,deconick2004}

\bea\label{model}
\fl i\frac{\partial\psi_{j}}{\partial t}+\gamma(x)\frac{\partial^{2}\psi_{j}}{\partial x^{2}}+2\sum^{2}_{l=1}g_{jl}(x)|\psi_{l}|^{2}\psi_{j}
+V(x,t)\psi_{j}-\sum^{2}_{l=1,l\neq j}\sigma\psi_{l}=0,\quad j=1,2,
\eea
where $\psi_{j}$ is the macroscopic wave function of the $j^{th}$ component in the context of BEC and it represents the complex electric field envelope of the $j^{th}$
component in optical setting. In equation (\ref{model}), $t$ is the time coordinate and $x$ is the spatial co-ordinate in the  BEC setting, while in optics
respectively they are the longitudinal and transverse coordinates. The dispersion parameter $\gamma(x)=\frac{\hbar}{2 m(x)}$, where $m$ is the effective
mass, $\sigma$ is the Rabi coupling coefficient responsible for the interaction of two or more different hyper fine states atoms, $g_{jl}(x)$ are the
nonlinearity coefficients related to the $s$-wave scattering lengths $a_{jl}(x)$ through $g_{jl}(x)=\frac{4\pi \hbar^{2}a_{jl}(x)}{a_{B}}$,
where $a_{B}$ is the Bohr radius. Based on experimental results pertaining to two-component $^{87}$Rb BECs \cite{hamner2011,middelkamp2011,yan2011},
we can assume that the scattering length ratios are almost equal to one; and also they can be tuned through Feshbach resonance, as it was demonstrated in
experimental works \cite{thalhammer2008,papp2008}. Hence, without loss of generality, we consider equal interaction strengths, \textit{i.e.,}
$g_{jl}(x)=g(x)$ in equation (\ref{model}). Here  $V(x,t)$ represents external spatially varying trapping potential .
\subsection{Transformation to the Manakov system}

Here our aim is to transform equation (\ref{model}) to an integrable nonlinear evolution equation with constant coefficients, namely canonical Manakov system. For this purpose, we first perform the rotational transformation
\bea \label{unitary}
\left(\begin{array}{l}
	\psi_{1}(x,t)\\
	\psi_{2}(x,t)\\
\end{array} \right)=\left(\begin{array}{cc}
	\cos(\sigma t) & -i\sin(\sigma t)\\
	-i\sin(\sigma t)& \cos(\sigma t)\\
\end{array} \right)~~\left(\begin{array}{l}
	\phi_{1}(x,t)\\
	\phi_{2}(x,t)\\
\end{array} \right),
\eea
in equation (\ref{model}). This results in the following inhomogeneous 2-CGP system
\bea\label{cgpes}
i\frac{\partial\phi_{j}}{\partial t}+\gamma(x)\frac{\partial^{2}\phi_{j}}{\partial x^{2}}+2g(x)\sum^{2}_{l=1}|\phi_{l}|^{2}\phi_{j}+V(x,t)\phi_{j}=0,\qquad j=1,2.
\eea
The above equation (\ref{cgpes}) can then be transformed to the celebrated Manakov system with the canonical form \cite{manakov1974}
\bea\label{cnls}
i\frac{\partial q_{j}}{\partial T}+\frac{\partial^{2}q_{j}}{\partial Y^{2}}+2\sum^{2}_{l=1}|q_{l}|^{2}q_{j}=0,\qquad j=1,2,
\eea
by employing the similarity transformation
\bea \label{trans} \phi_{j}(x,t)=N(x)~q_{j}(T,Y), \quad j=1,2,\eea
where $Y=Y(x)$ and $T=t$ are self-similar variables.

The real amplitude $N(x)$ in (\ref{trans}) satisfies the following set of coupled ordinary differential equations
\bes
\bea \label{YN}
\gamma(x) \frac{d^{2}N(x)}{dx^{2}}+V(x,t)N(x)=0,\\
\label{Y}
2\frac{dN(x)}{dx}\frac{dY(x)}{dx}+N(x)\frac{d^{2}Y(x)}{dx^{2}}=0,\eea
where the similarity variable $Y(x)$ is defined in terms of amplitude as
\bea \label{Yg}
Y(x)=\int \frac{dx}{[N(x)]^{2}}~.
\eea
The dispersion parameter $\gamma(x)$ and nonlinearity function $g(x)$ are related to $N(x)$ respectively as
\bea
\gamma(x)=[ N(x)]^{4},~~g(x)= [N(x)]^{-2}.
\eea \label{intcon}
\ees
Here we confine ourselves to spatially dependent potentials only i.e.,$V(x,t)\equiv V(x)$ in view of experiments on BEC \cite{prl2003}. The conditions on $g(x), \gamma(x)$ and $Y(x)$ given by equation (6) can be viewed as the integrability condition for the inhomogeneous 2-CGP system (\ref{cgpes}) and hence that of (\ref{model}). The above equations clearly show that all the variable coefficients can be expressed in terms of $N(x)$. Thus for a given form of trapping potential, one can determine the form of $N(x)$ from (\ref{YN}) which in turn results in the required dispersion parameter function and necessary nonlinearity management for obtaining exact nonlinear waves in equation (\ref{model}).

In this work, we consider two types of trapping potentials namely optical lattice potential \cite{bronski2001a,bronski2001b,Morsch2006}
and parabolic cylinder potential \cite{taylor1992} for illustrative purpose.
\begin{enumerate}
	\item Periodic Potentials (optical lattice potential)

First we are interested in the lattice potential having the general form
\bea \label{OL-poten}
V(x)=\gamma(x)(\lambda-\ell(\ell+1)m \sn^{2}(x,m)),
\eea
which is similar to the Lam\'{e} potential with $\lambda$ being the eigenvalue of the Lam\'{e} equation \cite{arscott1964,kanna2016},
$\ell$ is the order of the Lam\'{e} polynomial and $\sn(x,m)$ denotes the Jacobian ellipticity sine function with elliptic modulus $m$, $0\leq m\leq1$. In this work, we consider first order Lam\'{e} polynomials  $\sn(x,m), \cn(x,m)$ and $\dn(x,m)$ as eigenfunctions with eigenvalues $(\lambda =)~ m+1, 1$ and $m$, respectively. The potential is sinusoidal for the case  $m=0$ and thus describes an optical lattice potential. For $m=1$, it results in the hyperbolic potential and the potential becomes doubly periodic if $m$ lies between 0 to 1. Such potentials expressed in terms $\sn^{2}(x,m)$ function can be well approximated by a periodic lattice potential created from four interacting laser beams \cite{brazhnyi2004}. Experimentally, one can clearly observe that the matter wave dynamics can be changed and also be enhanced considerably when an optical lattice is imposed on the condensate \cite{anderson1998}.\\
	
	\begin{figure}[t]
\centering\includegraphics[width=0.95\linewidth]{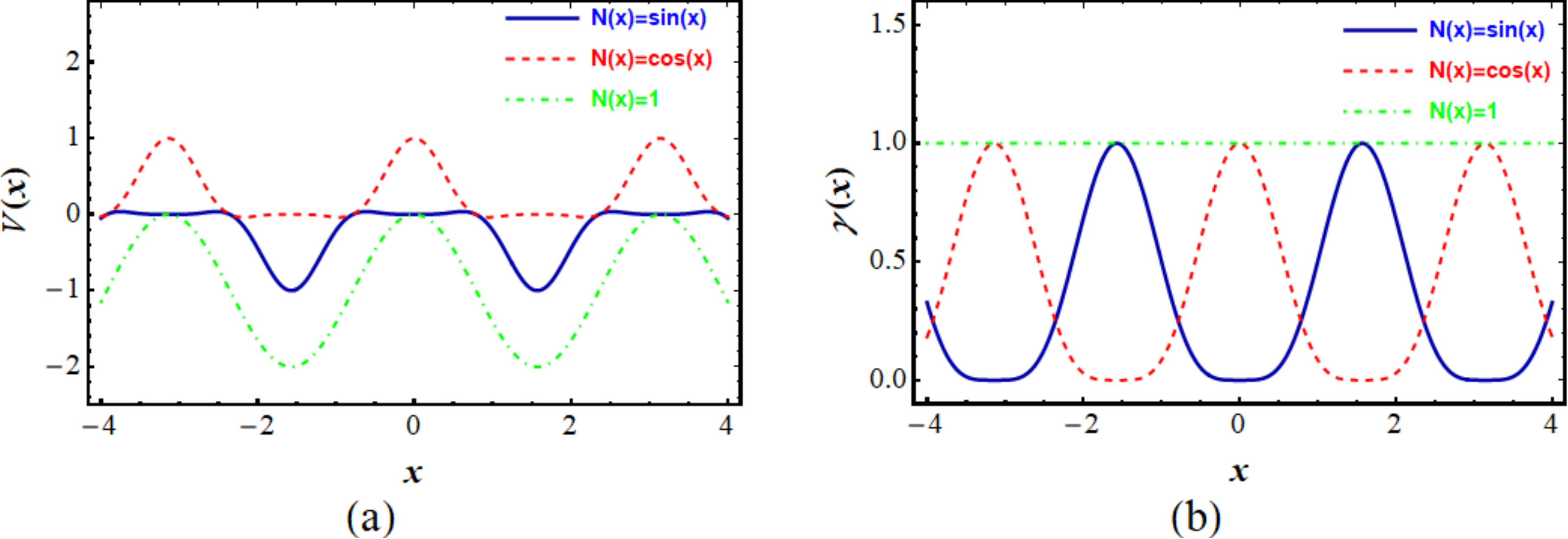}\\
\includegraphics[width=0.95\linewidth]{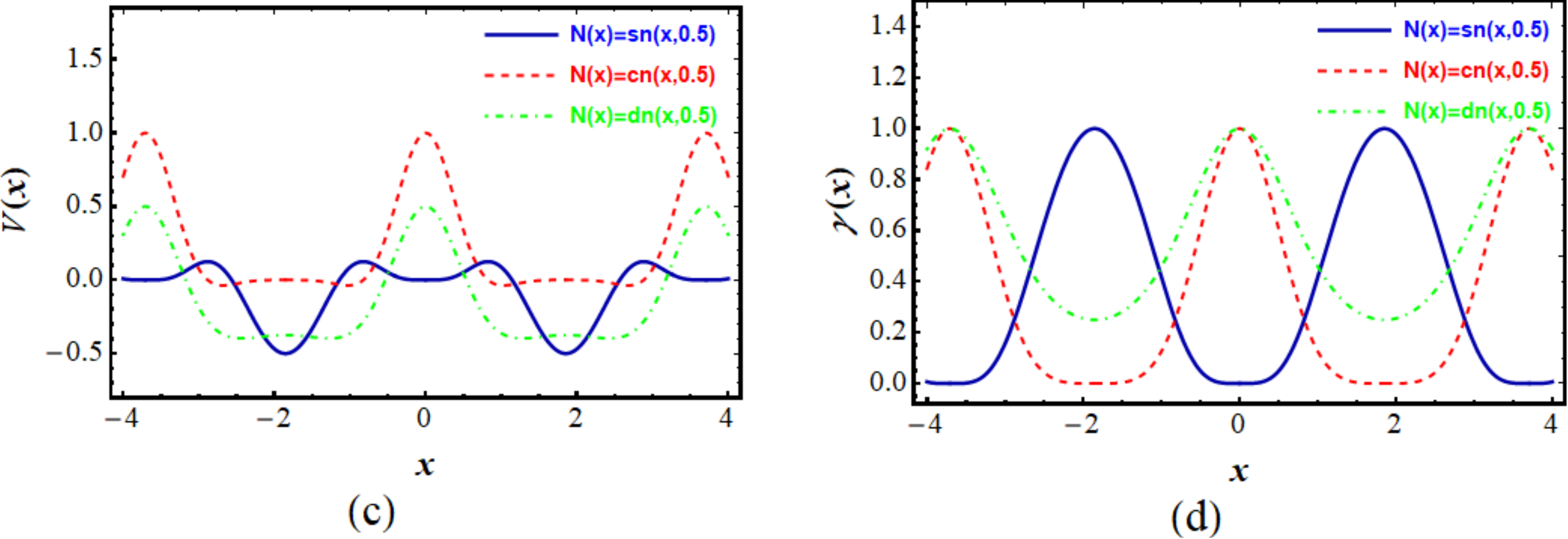}\\
\includegraphics[width=0.95\linewidth]{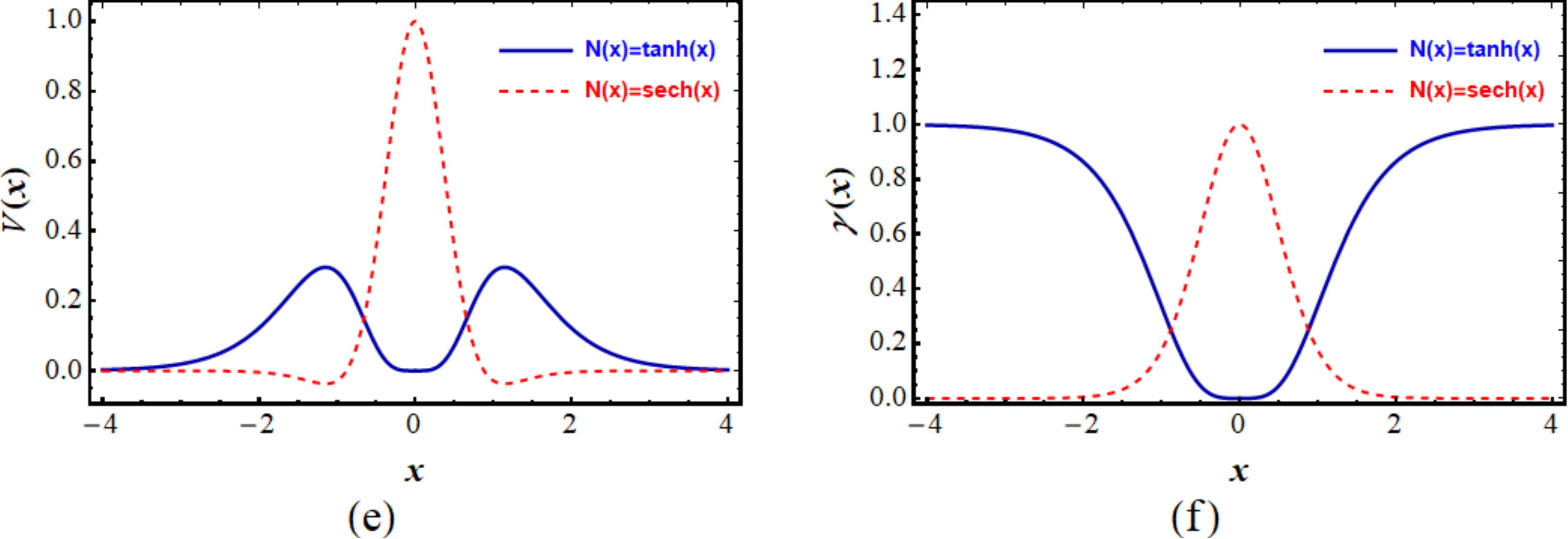}\\
\caption{The top, middle and bottom panels show profiles of trapping potential $V(x)$ and dispersion parameter function $\gamma(x)$  for differen choices of $N(x)$ resulting for $m=0,0.5,1,$ respectively.}
\label{fig1}
	\end{figure}

	The profiles of optical lattice potentials for $\ell=1$ and the resulting $\gamma(x)$ are shown respectively in the left and
	right panels of Fig.~\ref{fig1}.

	\item Next, we consider the parabolic cylinder potential having a fairly general form:
	\bea \label{PC-poten} V(x)=\gamma(x) (ax^2+b),\eea
	where $a$ and $b$ are real constants and are chosen as $\frac{-1}{4}$ and $n+\frac{1}{2}$ respectively, $n$ is the quantum model parameter which is a non-negative integer. Figs.~\ref{fig2}(a-b) represent the form of the parabolic cylinder potential $V(x)$ and the corresponding dispersion parameter function $\gamma(x)$ for
	$n=0,1,2,$ respectively. The functional forms of $N(x)$ and $\gamma(x)$ corresponding to  the above two potentials are listed in Table-\ref{tabel1}.
\end{enumerate}

%%%%%%%%%%%%%%%%%%%%%%%%%%%%%%%%%%%%%
\vspace{0.5cm}
\begin{figure}[t]
\centering\includegraphics[width=0.95\linewidth]{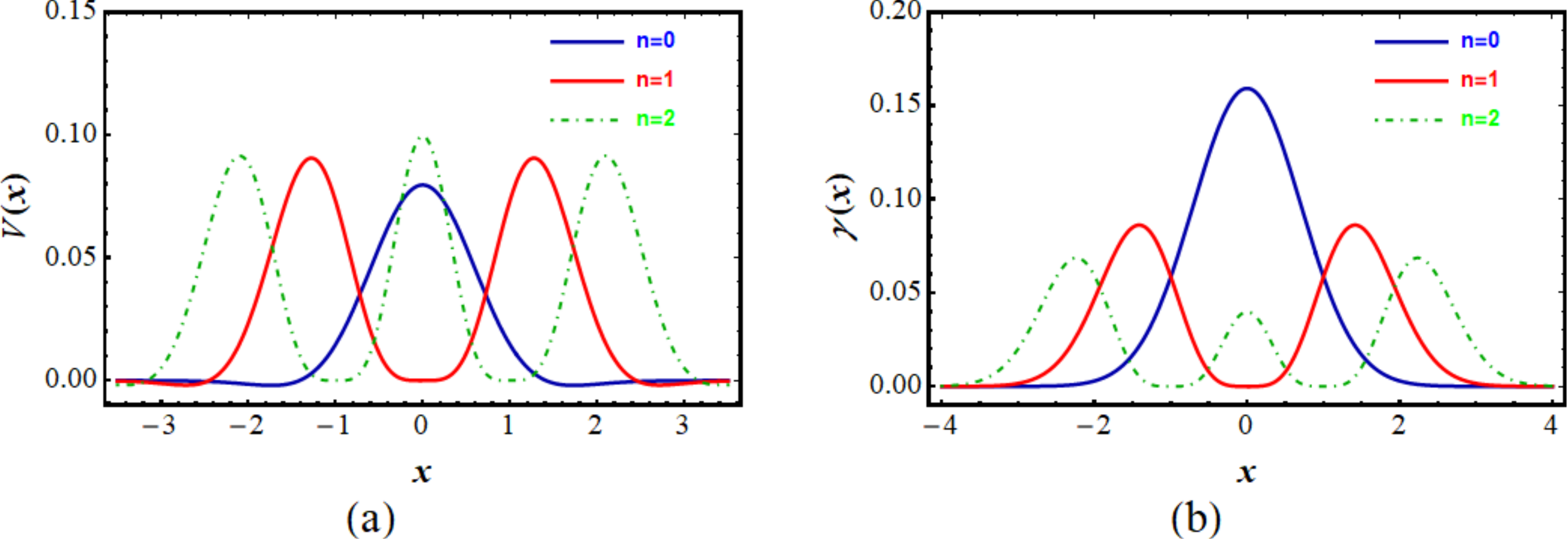}
	\caption{ Profiles of trapping potential $V(x)$ and the dispersion parameter function $\gamma(x)$ for different values of model parameter $n$.}
	\label{fig2}
\end{figure}
\begin{table}[h]
	\centering
	\caption{Forms of potentials $V(x)$ satisfying the integrability condition (equation (6)) along with their corresponding amplitudes $N(x)$, and dispersion parameter function $\gamma(x)$.}
	\begin{tabular}{|c|l|l|l|}
		\hline
		S.No.&~~~~~~~~~$V(x,t)$~~~~~~ & ~~~$N(x)$~~~ & ~~~$\gamma(x)$~~~\\
		&&&\\ \hline
		1.&$\gamma(x)((m+1)-2m~\sn^{2}(x,m))$&$\sn(x,m)$& $\sn^{4}(x,m)$\\
		&$\gamma(x)(1-2m~\sn^{2}(x,m))$&$\cn(x,m)$& $\cn^{4}(x,m)$\\
		&$\gamma(x)(m-2m~\sn^{2}(x,m))$&$\dn(x,m)$&$\dn^{4}(x,m)$\\
		\hline
		2.&$\gamma(x) (-\frac{1}{4}x^2+\frac{1}{2}+n)$& $k(c_{1}D_{n}(x)+$& $(k(c_{1}D_{n}(x)+$\\
		&&$c_{2}D_{-(n+1)}(ix))$&$c_{2}D_{-(n+1)}(ix)))^{4}$\\
		\hline
	\end{tabular}\label{tabel1}
\end{table}

\subsection{General rogue wave solutions}

The general RW (semi-rational) solution of the canonical Manakov system (\ref{cnls}) has been obtained by the DT method in Ref.~\cite{baronio2012} in a semi-rational form as
\bea
\label{tcrw}
\left(\begin{array}{l}
	q_{1}(Y,T)\\

	q_{2}(Y,T)\\
\end{array}\right)=\left[
\frac{L}{B}\left(\begin{array}{l}
	a_{1}\\
	a_{2}\\
\end{array}\right)+\frac{M}{B}\left(\begin{array}{l}
	~~a_{2}\\
	-a_{1}\\
\end{array}\right)\right] e^{2i\omega T},
\eea
where
$L=\frac{3}{2}-8\omega^{2}T^{2}-2a^{2}Y^{2}+8i\omega T+|f|^{2}e^{2aY},~M=4f(aY-2i\omega T-\frac{1}{2})e^{aY+i\omega T},
~B=\frac{1}{2}+8\omega^{2}T^{2}+2a^{2}Y^{2}+|f|^{2} e^{2aY},~\omega=a^{2},~a=\sqrt{a_{1}^{2}+a_{2}^{2}}$. Here $a_{1}$
and $a_{2}$ are real arbitrary parameters, $f$ is an  arbitrary complex constant. This solution features standard RW for the choice $a_{1}=a_{2}\neq0, f=0$ and interesting boomeronic solitons (comprising of both bright and dark parts)
accompanied by RWs for the parametric choice $a_{1}=a_{2}=f\neq0$.
\subsubsection{Constant coefficients 2-CGP system with Rabi coupling}

\begin{figure}[t]
\centering\includegraphics[width=0.95\linewidth]{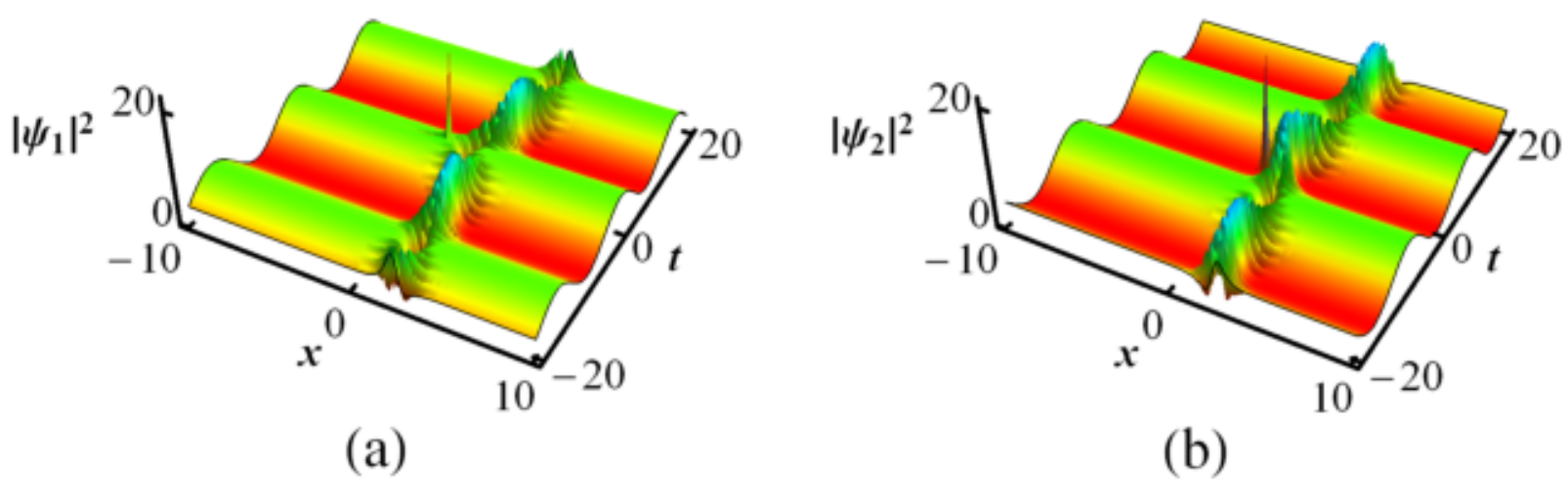}\\
\includegraphics[width=0.19\linewidth]{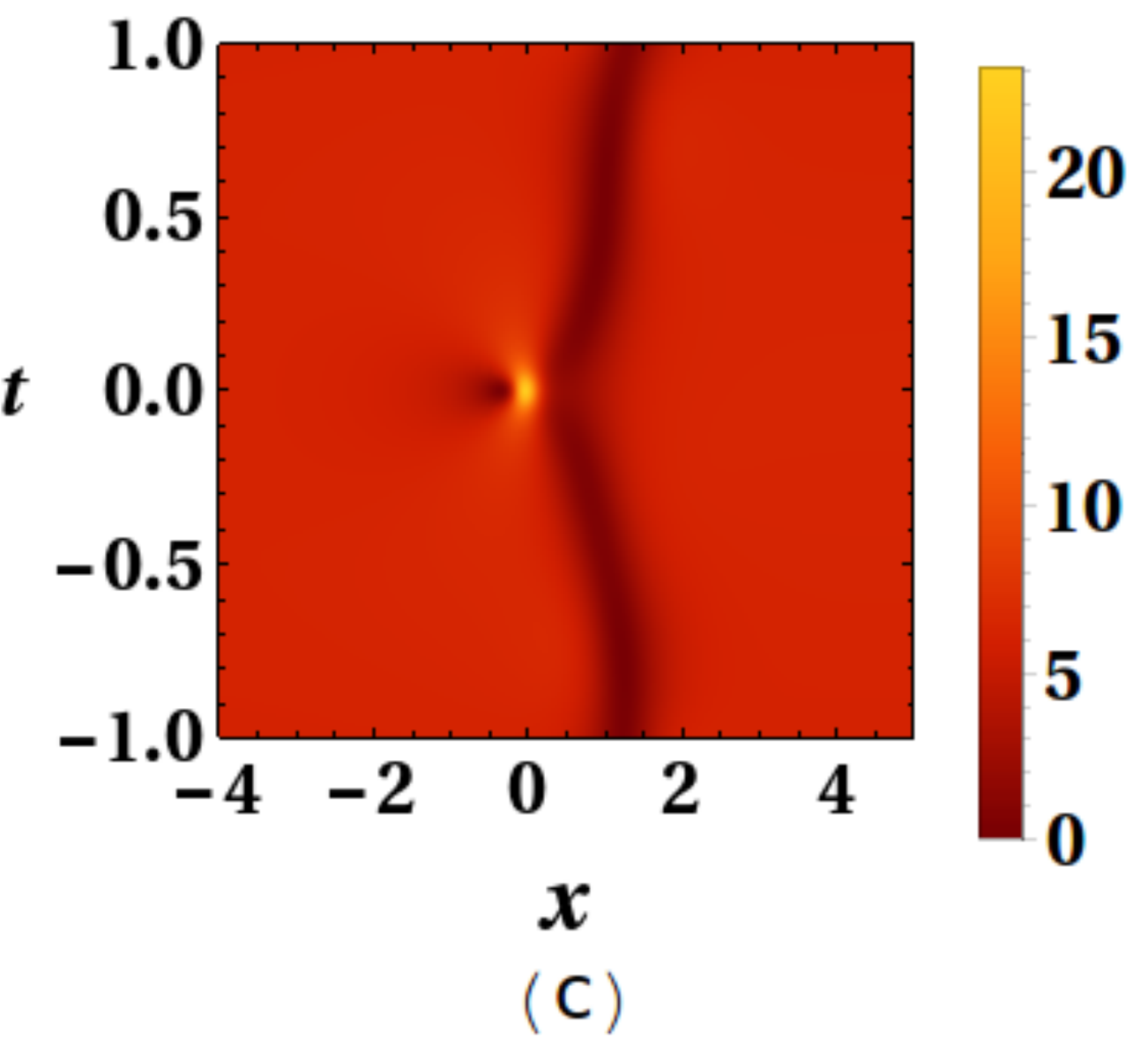}
\includegraphics[width=0.19\linewidth]{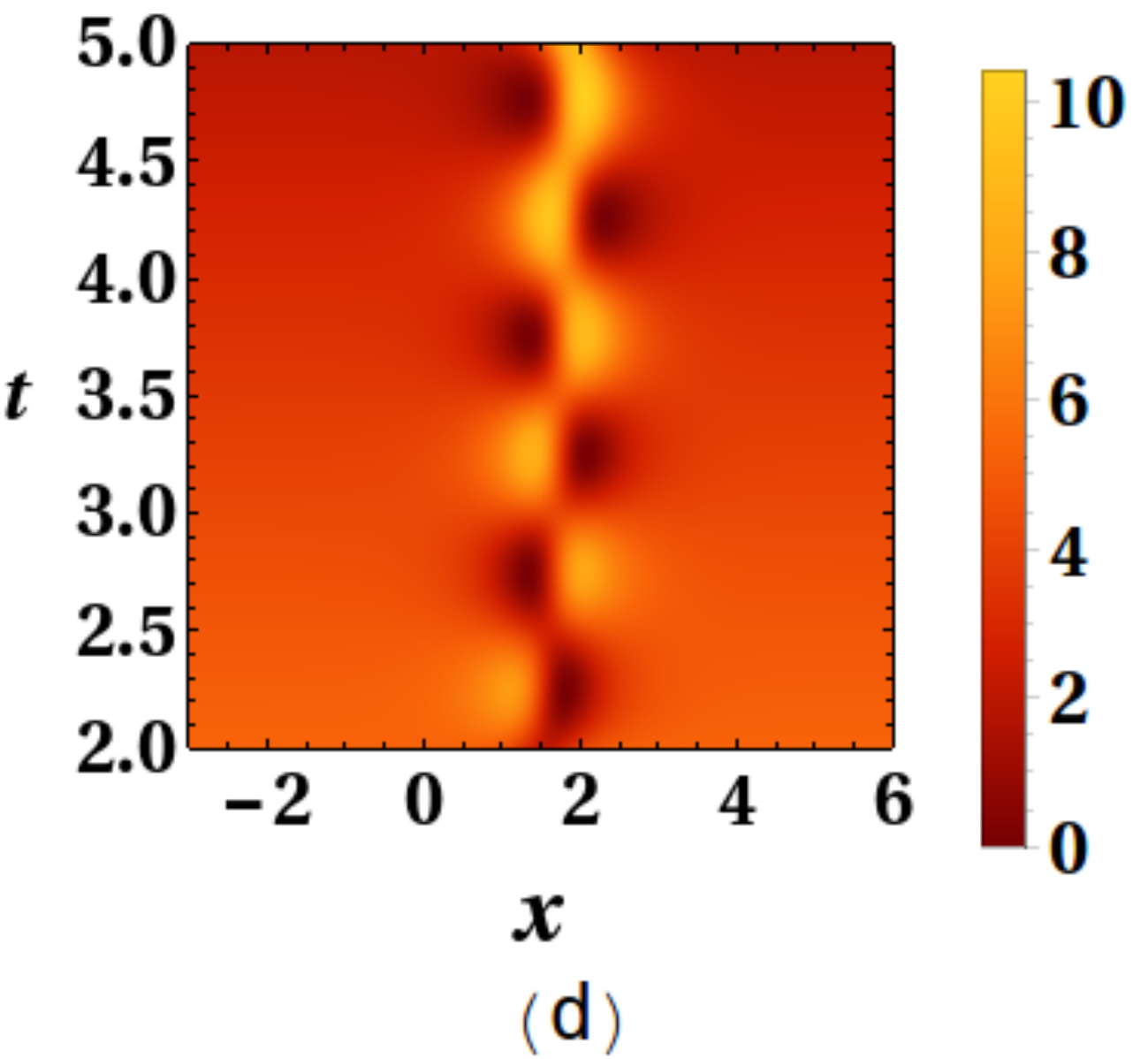}
\includegraphics[width=0.2\linewidth]{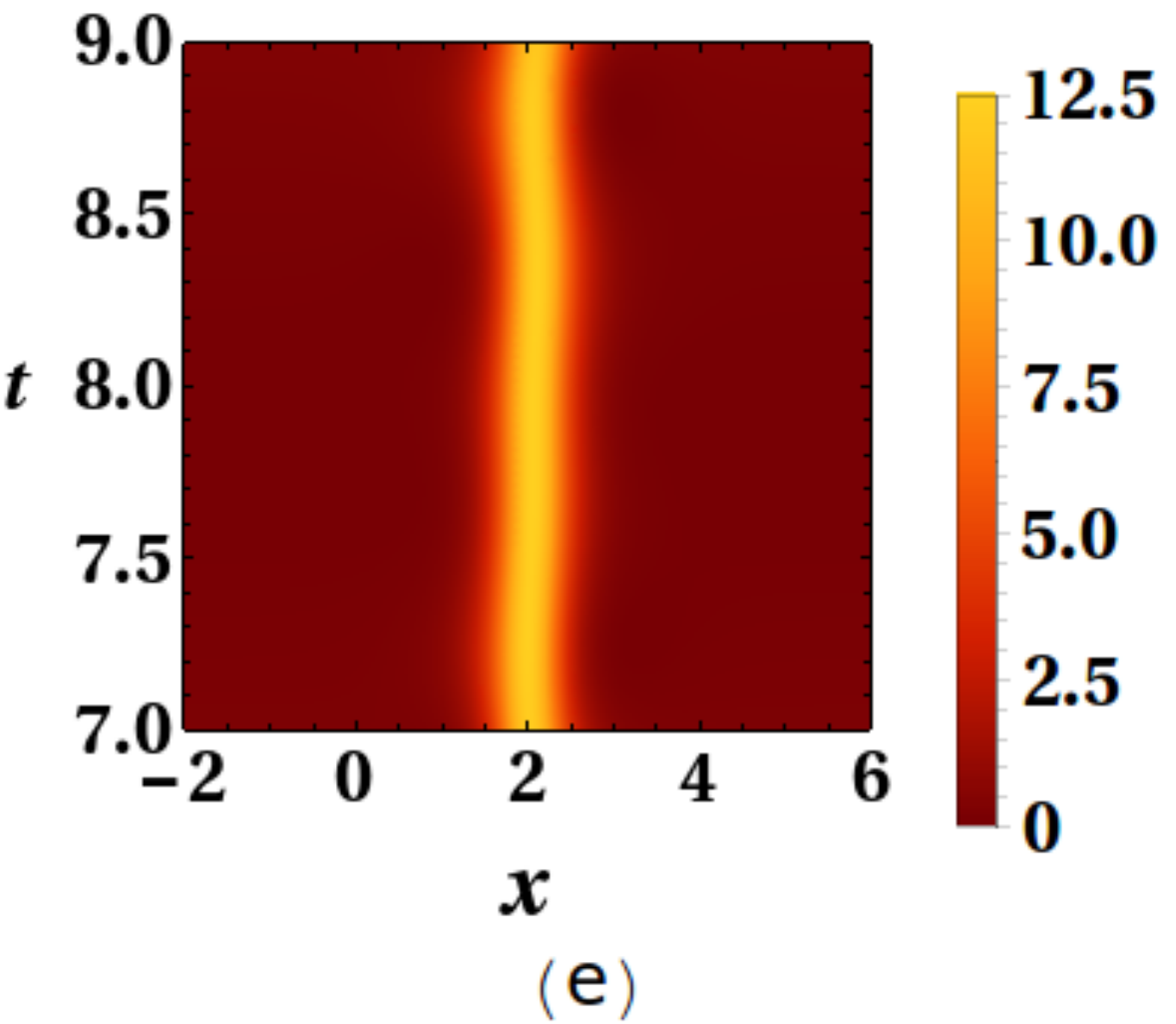}
\includegraphics[width=0.19\linewidth]{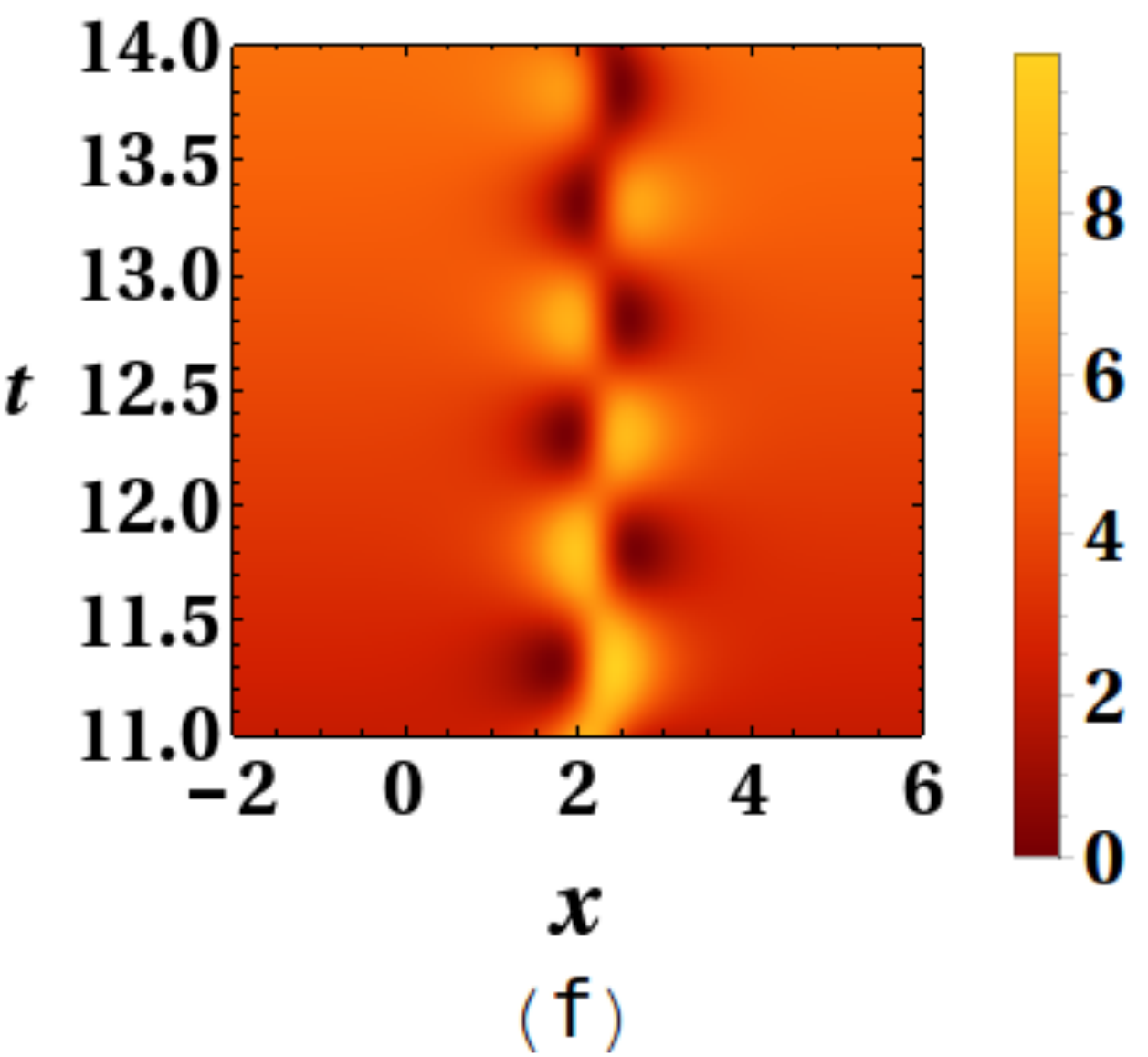}
\includegraphics[width=0.19\linewidth]{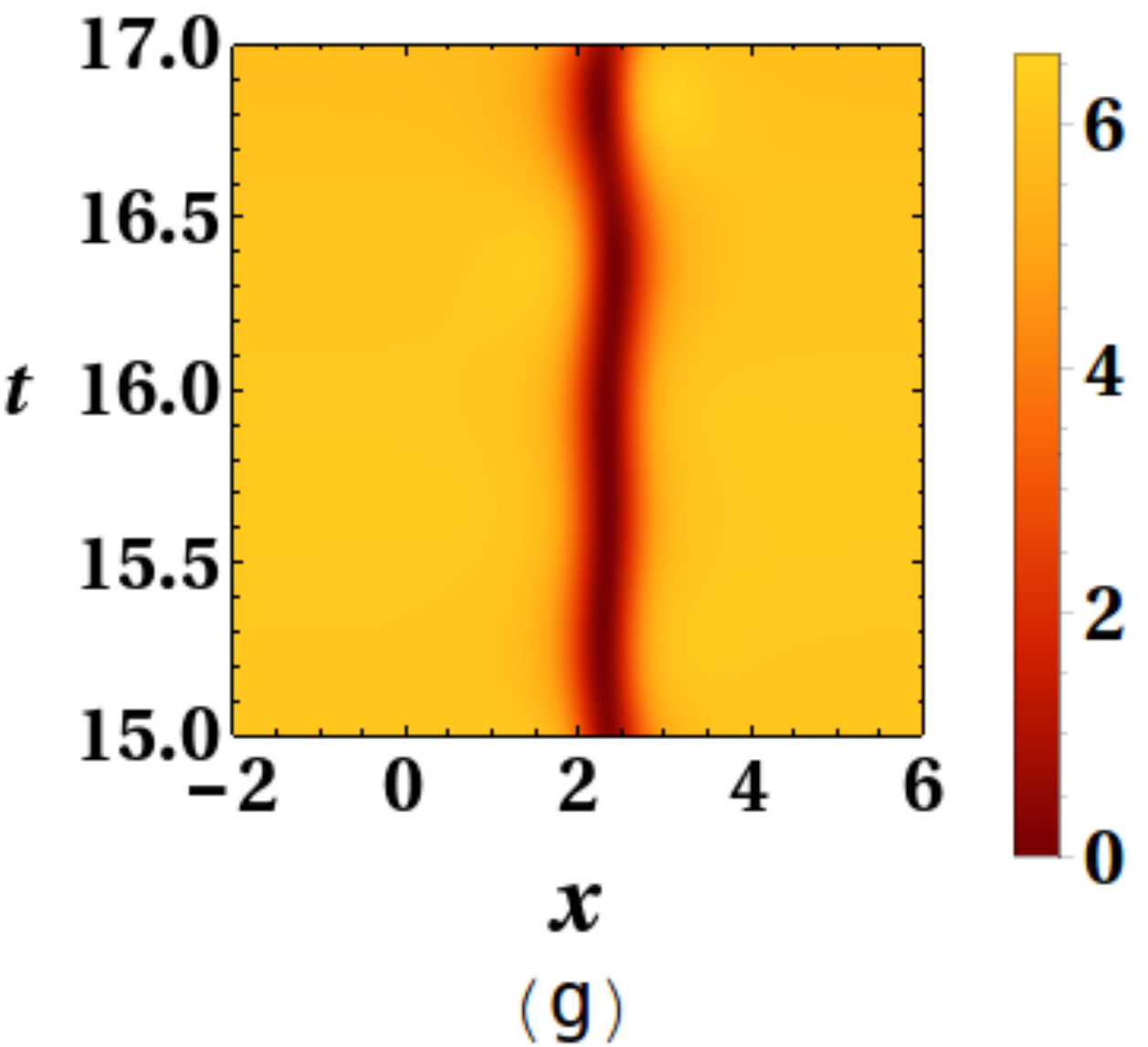}\\
\includegraphics[width=0.195\linewidth]{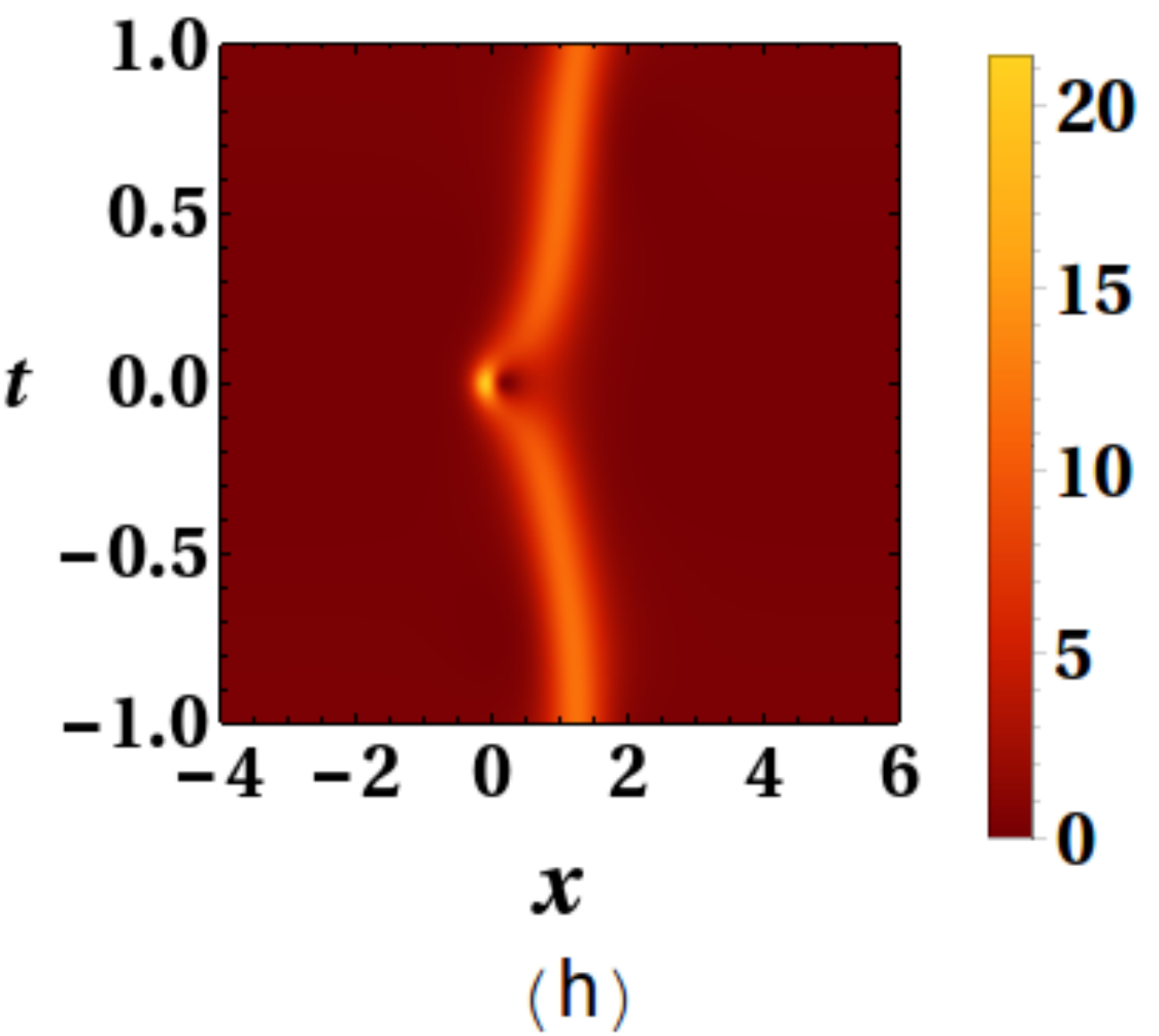}
\includegraphics[width=0.195\linewidth]{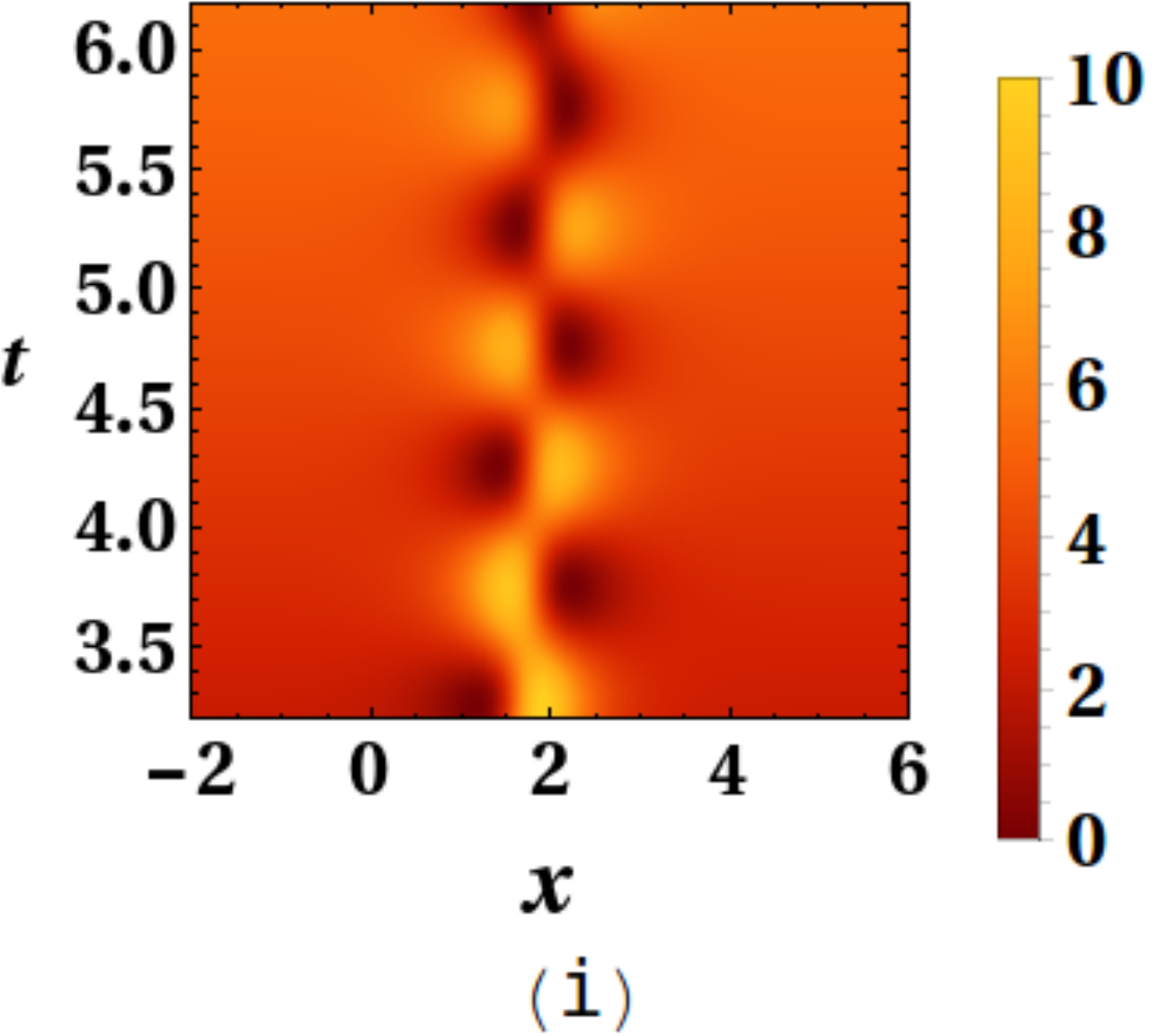}
\includegraphics[width=0.19\linewidth]{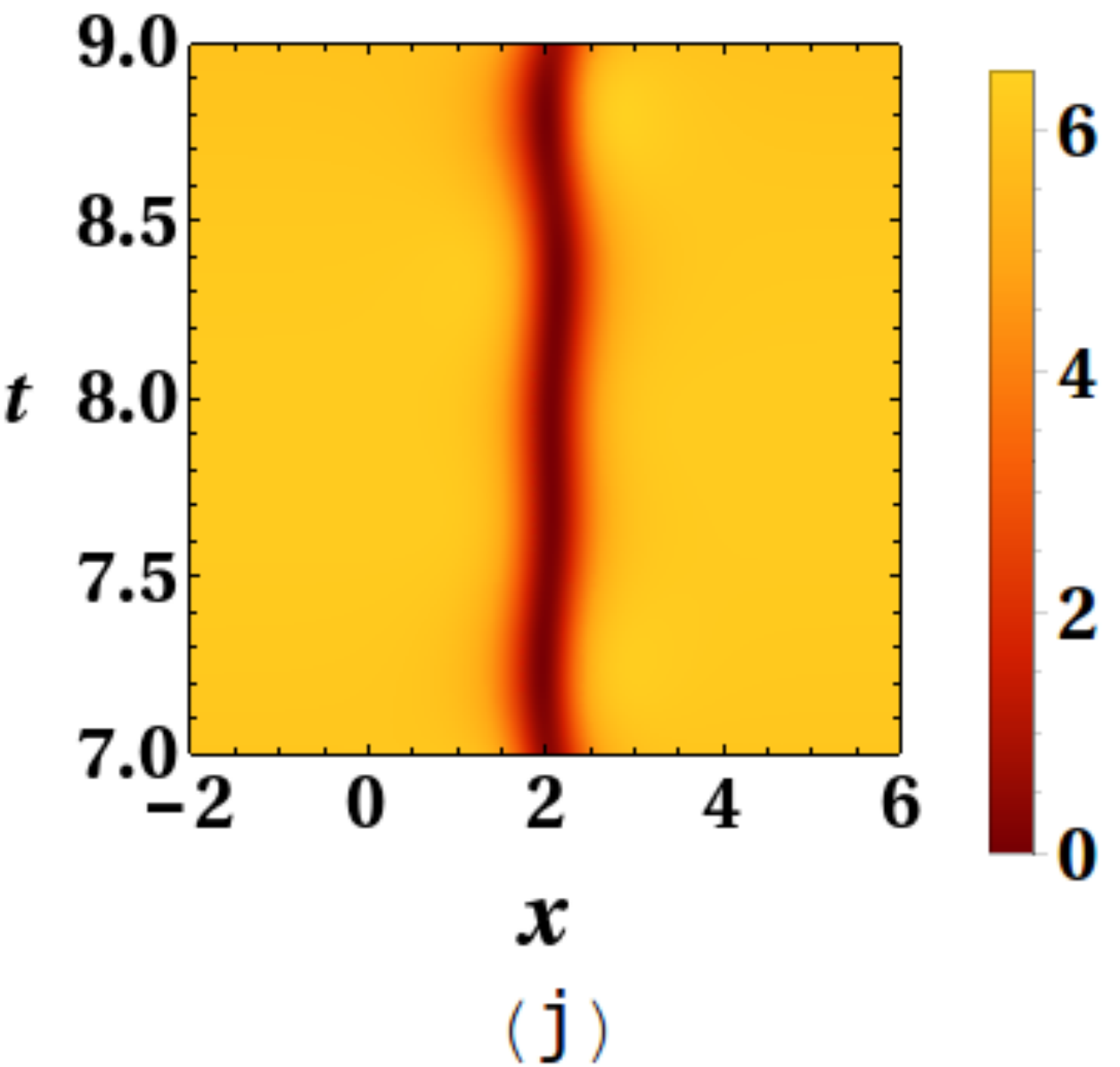}
\includegraphics[width=0.191\linewidth]{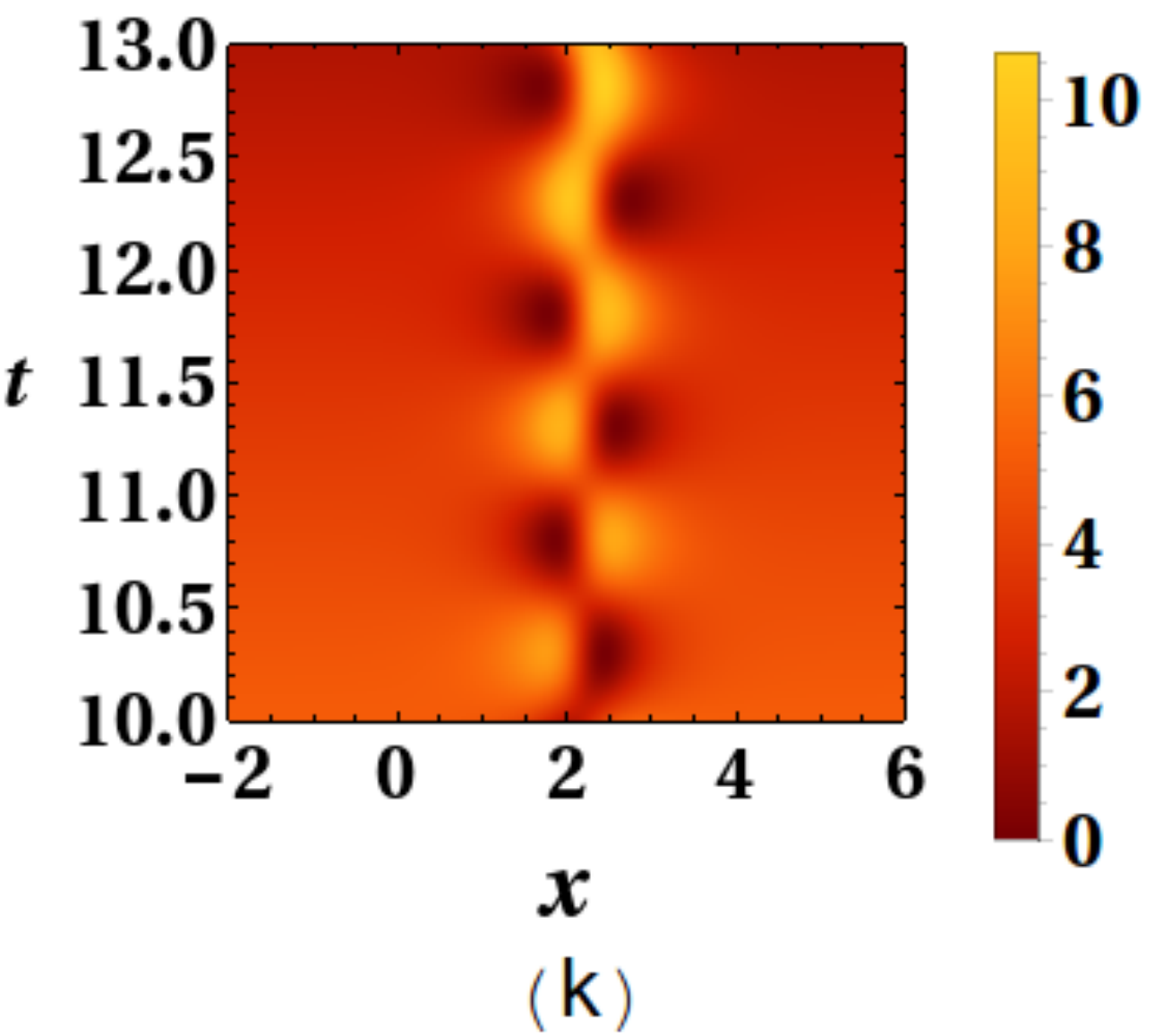}
\includegraphics[width=0.195\linewidth]{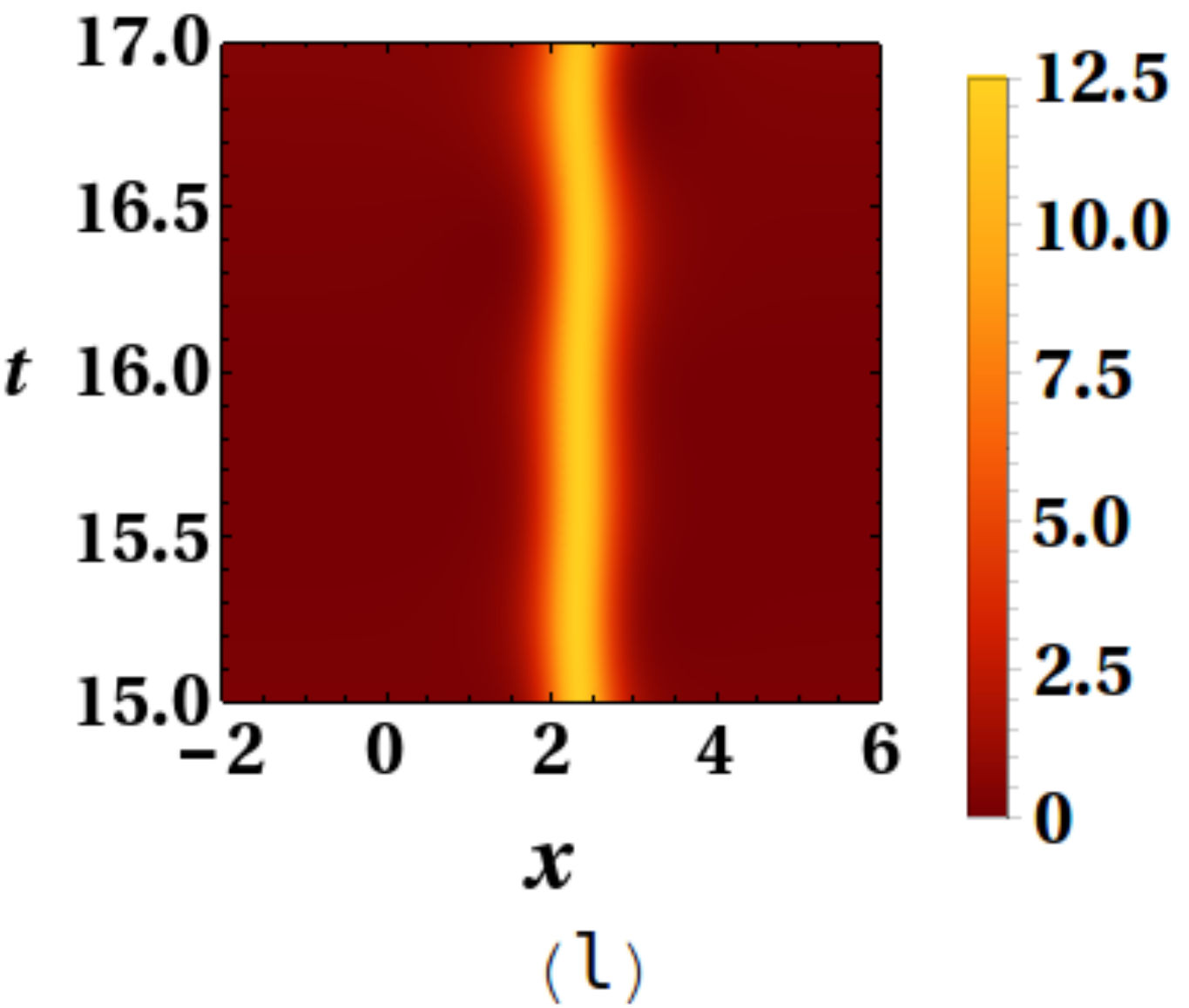}
	\caption{Exact rogue wave solutions in system (\ref{model}) with $a_{1}=2.5,~a_{2}=0,~f=0.8i,~\sigma=0.2,~g=\gamma=1$ and $V(x)=0$. Figs.~\ref{fig3}(a-b) are surface plots of the $\psi_{1}(x,t)$ and $\psi_{2}(x,t)$ components. Figs.~\ref{fig3}(c-l) shown that
		zoom-ins of panels (a) and (b).}
	\label{fig3}
\end{figure}

It is of relevance to briefly discuss the important results on Rabi coupled two-component GP system with constant coefficients before addressing the variable coefficient system. For this purpose we write down the corresponding solution for this case with $\sigma\neq0$, $g$ and $\gamma$ being constants in the absence of trapping potential
\bes\bea
\psi_{1}(x,t)&=&\cos(\sigma t)~q_{1}(x,t)-i~\sin(\sigma t)~q_{2}(x,t),\\
\psi_{2}(x,t)&=&-i~\sin(\sigma t)~q_{1}(x,t)+\cos(\sigma t)~q_{2}(x,t).
\eea\ees
Figure \ref{fig3} shows a special dynamical feature of the Manakov system in the presence of Rabi coupling alone. Top panels of Fig.~\ref{fig3} shows that RW co-exists with oscillating boomeronic soliton even for the choice $a_{1}=f\neq0, a_{2}=0$ in both the components for which boomeronic solitokn does not exist in the standard Manakov system. A careful look into the plot shows that at the center ($(x,t)=(0,0)$) there exists RW and a dark soliton. Then during first cycle  ($t$=0 to 15) of oscillation that central dark soliton transforms into a boomeronic soliton ($t=2$ to $5$), which later becomes as completely bright soliton ($t=7$ to $9$) and finally attains the original dark soliton form with an intermediate appearance of boomeronic soliton in the range ($t=11$ to $14$). This first cycle of oscillation is illustrated in the middle panels of Fig.~\ref{fig3}. This process repeats periodically. On the other hand, the second component ($\psi_{2}$) displays a reverse dynamical behaviour as shown in the bottom panels of Fig.~\ref{fig3}. Here the central bright soliton transforms to a boomeronic soltion, followed by an intermediate dark soliton form and finally retains the original bright form at the end of first cycle.

\subsubsection{Rabi coupled spatially modulated 2-CGP system}

Next, we turn our attention to the variable coefficient 2-CGP system (\ref{model}) for different trapping potentials.
Now using the similarity transformation equation (\ref{trans}) and the rotational transformation equation (\ref{unitary}), we construct the RW
solution of the two-component inhomogeneous GP equation with Rabi coupling (\ref{model}) as
\bes \bea \label{solu1}
\fl \psi_{1}(x,t)=N(x)\left(\frac{L}{B}\left(a_{1}\cos(\sigma t)-ia_{2}\sin(\sigma t)\right)+
\frac{M}{B}\left(i a_{1}\sin(\sigma t)+a_{2}\cos(\sigma t)\right)\right)e^{2i\omega t},\qquad\quad\\
\fl \psi_{2}(x,t)=N(x)\left(\frac{L}{B}\left(-ia_{1}\sin(\sigma t)+a_{2}\cos(\sigma t)\right)-
\frac{M}{B}\left(a_{1}\cos(\sigma t)+ia_{2}\sin(\sigma t)\right)\right)e^{2i\omega t}.\qquad\quad
\eea\ees
Here the amplitude function $N(x)$ depends on the various choices of external trapping potential.\\
\subsubsection{Optical Lattice potential}

(i) First, we consider the optical lattice potential discussed in the preceding section with the following form
\bea
V(x)= ((m+1)-2 m~\sn^{2}(x,m))\sn^{4}(x,m).
\label{snpf}\eea
Then from equation (\ref{YN}), the amplitude parameter is found to be $N(x)=\sn(x,m)$  and the integrability condition (\ref{YN}) results in
$\mbox Y(x)=\frac{\left(x-E(\varphi,m)\sn(x,m)\right)-\dn(x,m)\cn(x,m)}{\sn(x,m)}$, where $E$ is the elliptic integral of
second kind which in turn gives the spatially varying nonlinearity coefficient function as $g(x)=\frac{1}{\sn^{2}(x,m)}$ and the dispersion parameter function $\gamma(x)=\sn^{4}(x,m)$.

\begin{figure}[h]
\centering\includegraphics[width=0.95\linewidth]{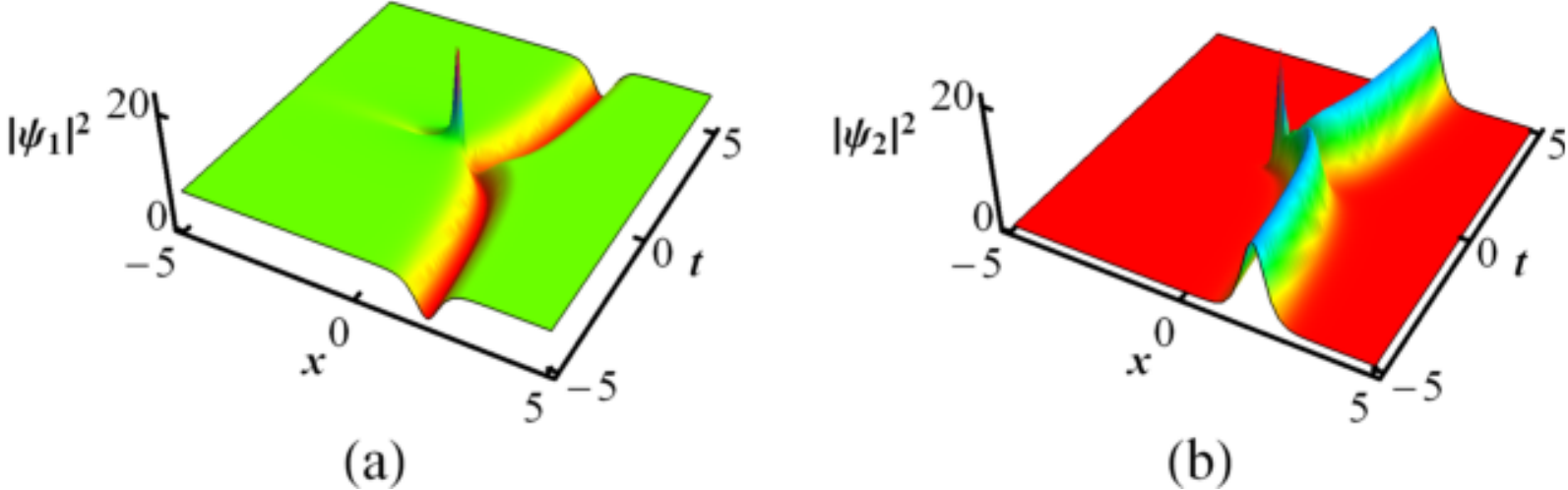}\\
\includegraphics[width=0.95\linewidth]{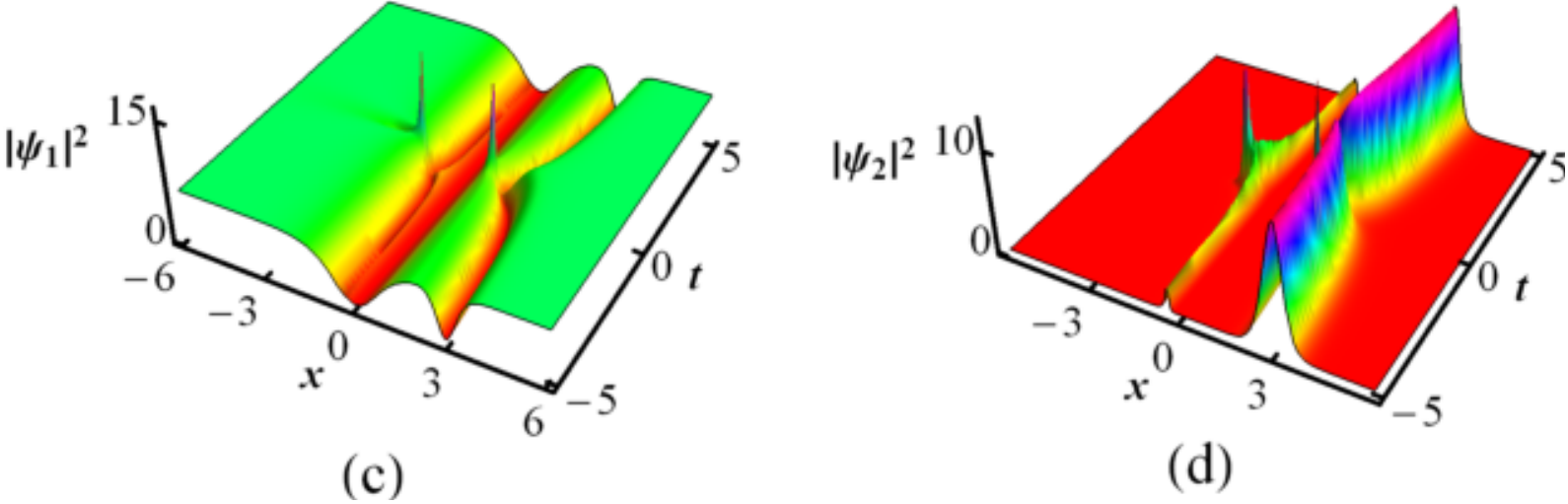}\\
\includegraphics[width=0.95\linewidth]{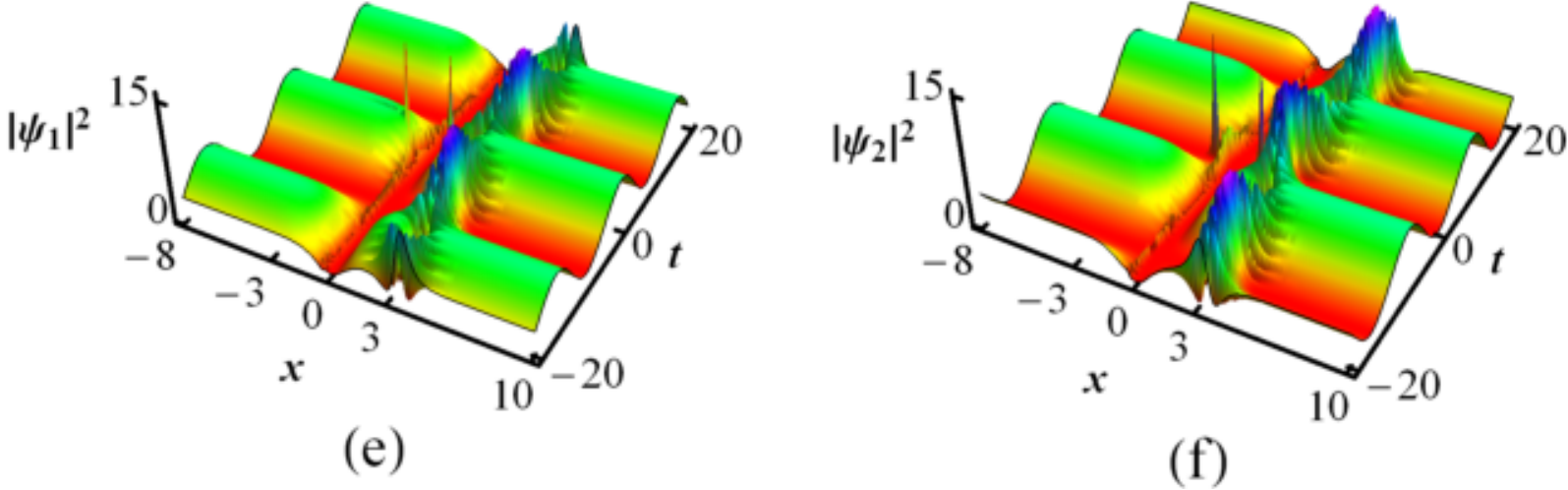}
	\caption{Exact RW solutions in the system (\ref{model}) for constant coefficients and in the absence of Rabi coupling (top panels); in the presence of variable coefficients without Rabi coupling (middle panels); and in the presence of both variable coefficients and Rabi coupling (bottom panels). The potential form is given by equation (\ref{tanhform}). The parameter values are chosen as $a_{1}=2.5,~a_{2}=0,~f=0.8,$ and $\sigma=0.045$.}
	\label{fig4}
\end{figure}

The nature of the vector RWs differs for different values of the modulus parameter $m$. Let us consider the modulus parameter $m$ to be unity. Ultimately $V(x)$ becomes
\bea \label{tanhform}
V(x)= 2(1-\tanh^{2}(x)) \tanh^{4}(x).
\eea
The standard Manakov soliton resulting for $\sigma=V(x)=0$ and $\gamma=g=1$ is shown in Figs.~\ref{fig4}(a-b) for comparison purpose. In the absence of Rabi coupling and trapping potential the resulting solution supports RW with dark soliton part and RW with bright soliton part in first and second components respectively (see Figs.~\ref{fig4}(a-b)). Note that due to the inclusion of trapping potential, new daughter RW with dark part and RW with bright part appears in first and second components respectively, there by shifting the original parent wave (RW combined with soliton) to the right with a suppression in RW amplitude and an enhancement in soliton part amplitude (see at region $x=3$). This is shown in Figs.~\ref{fig4}(c-d).

\begin{figure}[h]
\centering\includegraphics[width=0.95\linewidth]{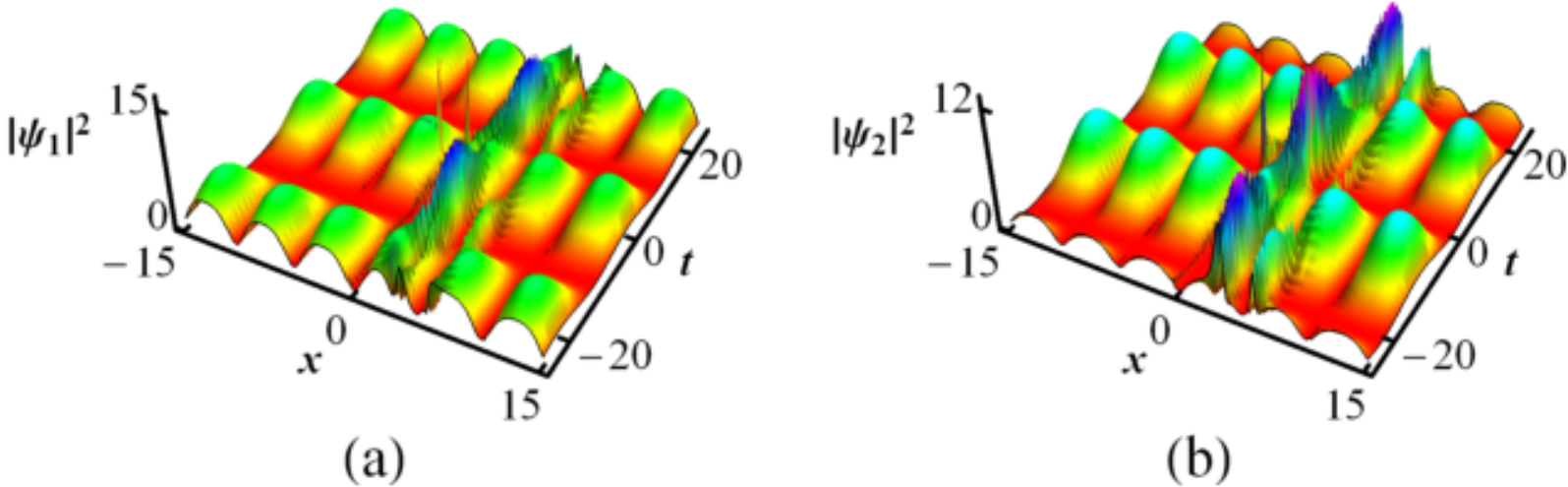}
\caption{Exact RW solutions of the system (\ref{model}) in the presence of both variable coefficients and Rabi coupling for the potential given by equation (\ref{snpf}). The parameter values $a_{1}=2.5,~a_{2}=0,~f=0.8,~m=0.9,$ and $\sigma=0.045$.}
	\label{fig5}
\end{figure}
Next we include Rabi coupling, say $\sigma=0.045$. Here the oscillating boomeronic behaviour discussed in Fig.~\ref{fig3}, is observed predominantly only in the shifted parent RW-soliton pair located at $x=3$ (see Fig.~\ref{fig4}). The newly created daughter RW along with dark soliton in an oscillating background remains almost unaffected. These are shown in Figs.~\ref{fig4}(e-f).

Further analysis of solution (11) for trapping potentials resulting for other values of $m~~(0<m<1)$ as given by (\ref{snpf}) is straight-forward. This choice will result in a structure resembling two dimensional optical lattice with modulations around the central region by RW and solitons which features the oscillating boomeronic soliton behaviour in that region. This is shown in Fig.~\ref{fig5}. \\

(ii) Next, we consider a different form for the optical lattice potential.
\bea \label{cnpf}
V(x)= (1-2 m\cn^{2}(x,m)) \cn^{4}(x,m),
\eea
which is depicted in Fig.~\ref{fig1}(e). The amplitude parameter is calculated from equation (\ref{YN}) as $N(x)=\cn(x,m)$  and the integrability condition yields $Y(x)=x+\frac{1}{(m-1)}\left(E(\varphi,m)-\frac{\sn(x,m)\dn(x,m)}{\cn(x,m)}\right)$. This determines the spatially varying nonlinearity coefficient function and the corresponding dispersion parameter function as
$g(x)=\frac{1}{\cn^{2}(x,m)}$ and $\gamma(x)=\cn^{4}(x,m)$ respectively.

\begin{figure}[t]
\centering\includegraphics[width=0.95\linewidth]{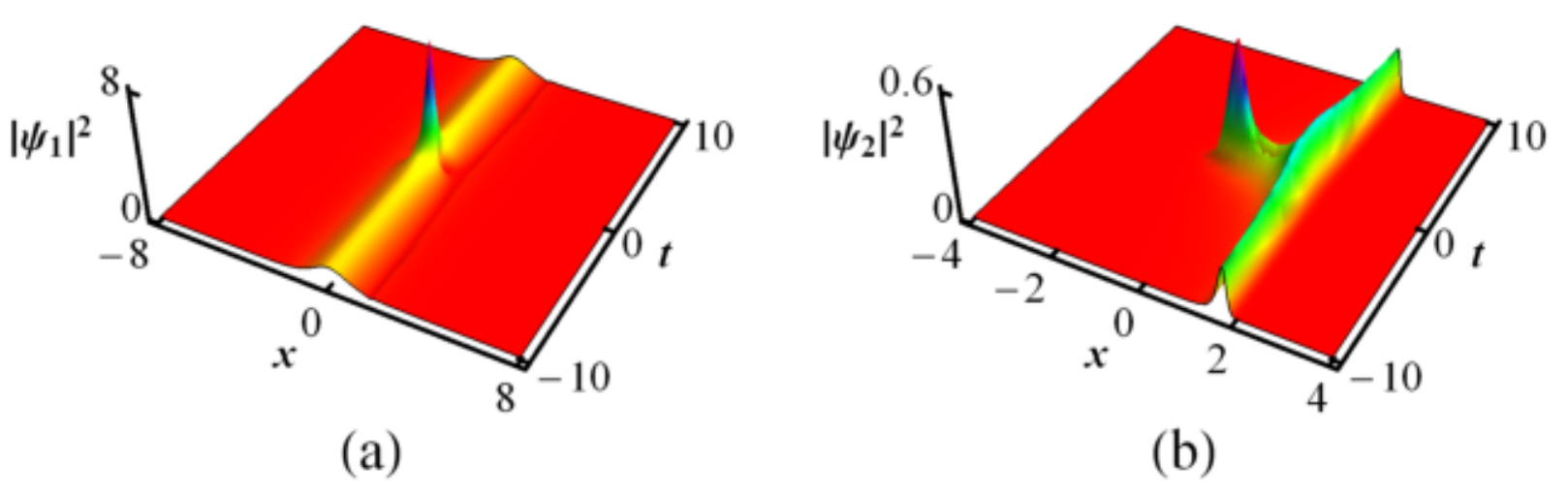}\\
\includegraphics[width=0.95\linewidth]{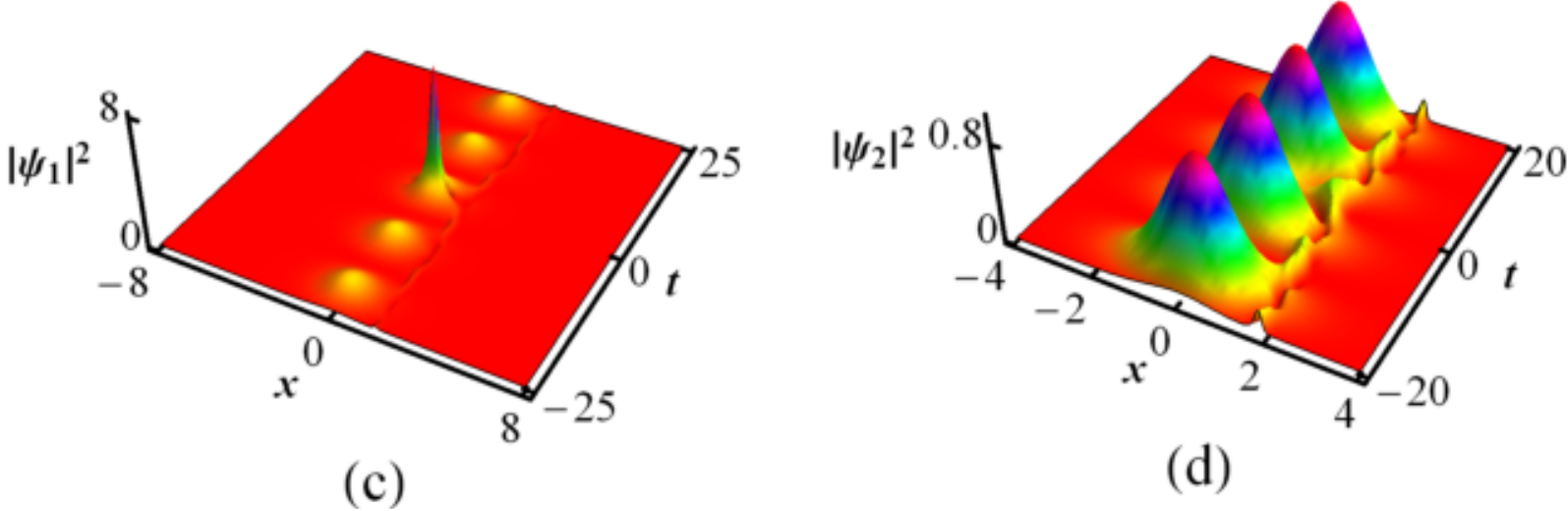}
\caption{Exact rogue wave solutions in the system (\ref{model}) for $\sigma=0$
	(top panels) and for $\sigma=0.2$ (bottom panels). The potential form is given by equation (\ref{sechform}). The parameters values are chosen as $a_{1}=1,~a_{2}=0,~f=0.3$.}
	\label{fig6}
\end{figure}

For the modulus parameter $m=1$, the potential (\ref{cnpf}) takes the form
\bea\label{sechform}
V(x)= (1-2 \sech^{2}(x))\sech^{4}(x).
\eea
To construct the solution of the spatially modulated two-component GP equation (\ref{model}) with and without Rabi coupling, we start with the semi-rational solution (11) of the Manakov system where first (second) component supporting dark (bright) soliton and a RW. In the absence of Rabi coupling $(\sigma=0)$, but in the presence of spatial modulation and external trapping potential in the first component ($\psi_{1}$) the background of RW disappears and a zero background (ZB) RW superimposes the bright soliton at ($(x,t)=(0,0)$) and in the $\psi_{2}$ component the ZBRW appears with bright soliton. Thus due to the trapping potential the original dark soliton disappears (see Fig.~\ref{fig9}(a)) in the first component and the bright soliton gets shifted from its position in the second component (see Fig.~\ref{fig6}(b)). By including Rabi coupling (say $\sigma=0.2$), we observe the creation of very low intensity dromion (localized structure in $x-t$ plane) trains which co-exist with RW at center in the $\psi_{1}$ component soliton as shown  in Fig.~\ref{fig6}(c). However, in the $\psi_{2}$ component we get dromion trains with significant intensity accompanied by oscillating bright soliton. Clearly from Fig.~\ref{fig6}(d) we note that when dromion attains maximum intensity, the oscillating bright soliton reaches minimum intensity. Thus Rabi coupling induces creation of dromion trains even in this (1+1) dimensional spatially modulated 2-CGP system.

\subsubsection{Parabolic Cylinder Potential}

Next, we focus our attention on another form of spatial modulation of the trapping potential strength namely parabolic cylinder potential (\ref{PC-poten})
\bea\label{pcpo}
V(x)= k^{4} D^{4}_{n}(x)\left(n+\frac{1}{2}-\frac{1}{4}x^{2}\right).
\eea
Interestingly, for this choice the integrability condition $Y(x)=\int [k D_{n}(x)]^{-2} dx$ turns out to be the standard Weber equation having the following solution \cite{Zwillinger1997} as
$ N(x)=k c_{1}D_{n}(x), n=0,1,2$. This in turn gives the spatially-varying nonlinearity coefficient function as $g(x)=(1/k^{2})D^{2}_{n}(x)$ and also the dispersion parameter function as $\gamma(x)=k^{4}D^{4}_{n}(x)$. Here $D_{n}(x)$ is the parabolic-cylinder function of order $n$ (control parameter), $c_{1}$ is an arbitrary integration constant, and $k=\frac{1}{\sqrt{2\pi}n!}$ is the normalization constant. Without loss of generality, one can choose the constant $c_{1}$ to be unity. Clearly, as $|x|\rightarrow\infty$ for any non-negative integer $n$, $D_{n}(x)$ is stable.

\begin{figure}[h]
\centering\includegraphics[width=0.44\linewidth]{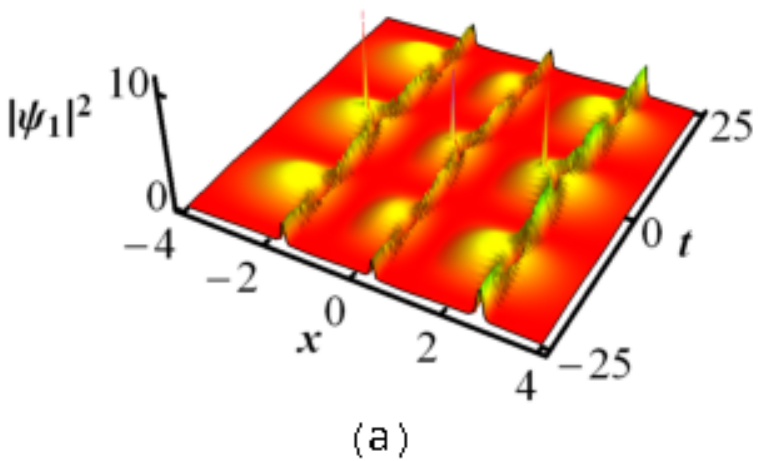}
~~~\includegraphics[width=0.34\linewidth]{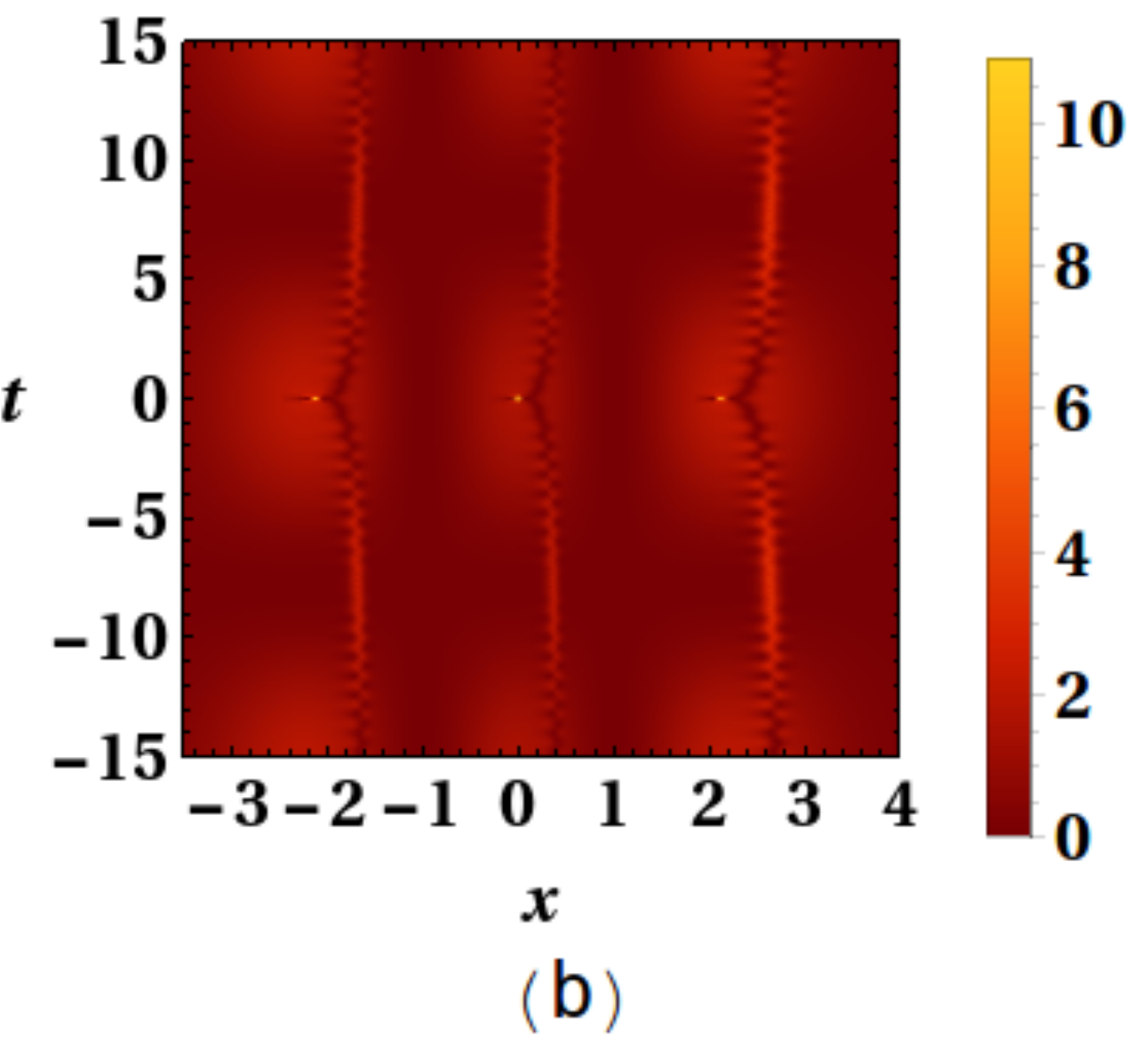}\\
\includegraphics[width=0.44\linewidth]{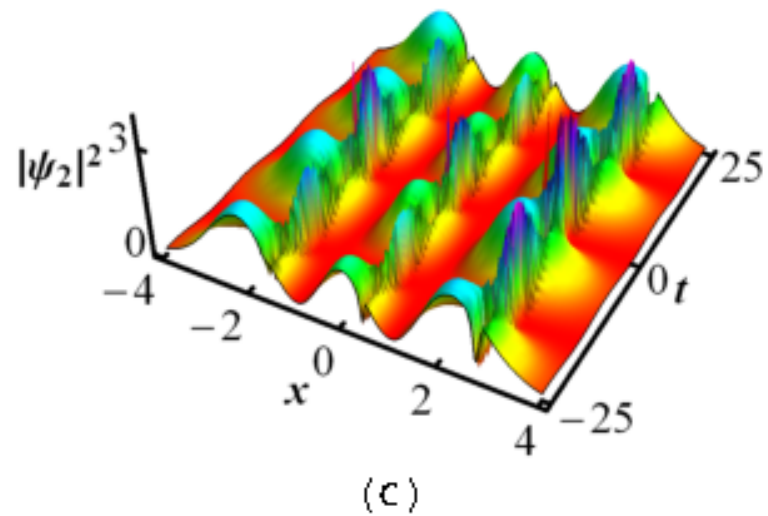}
	~~~\includegraphics[width=0.34\linewidth]{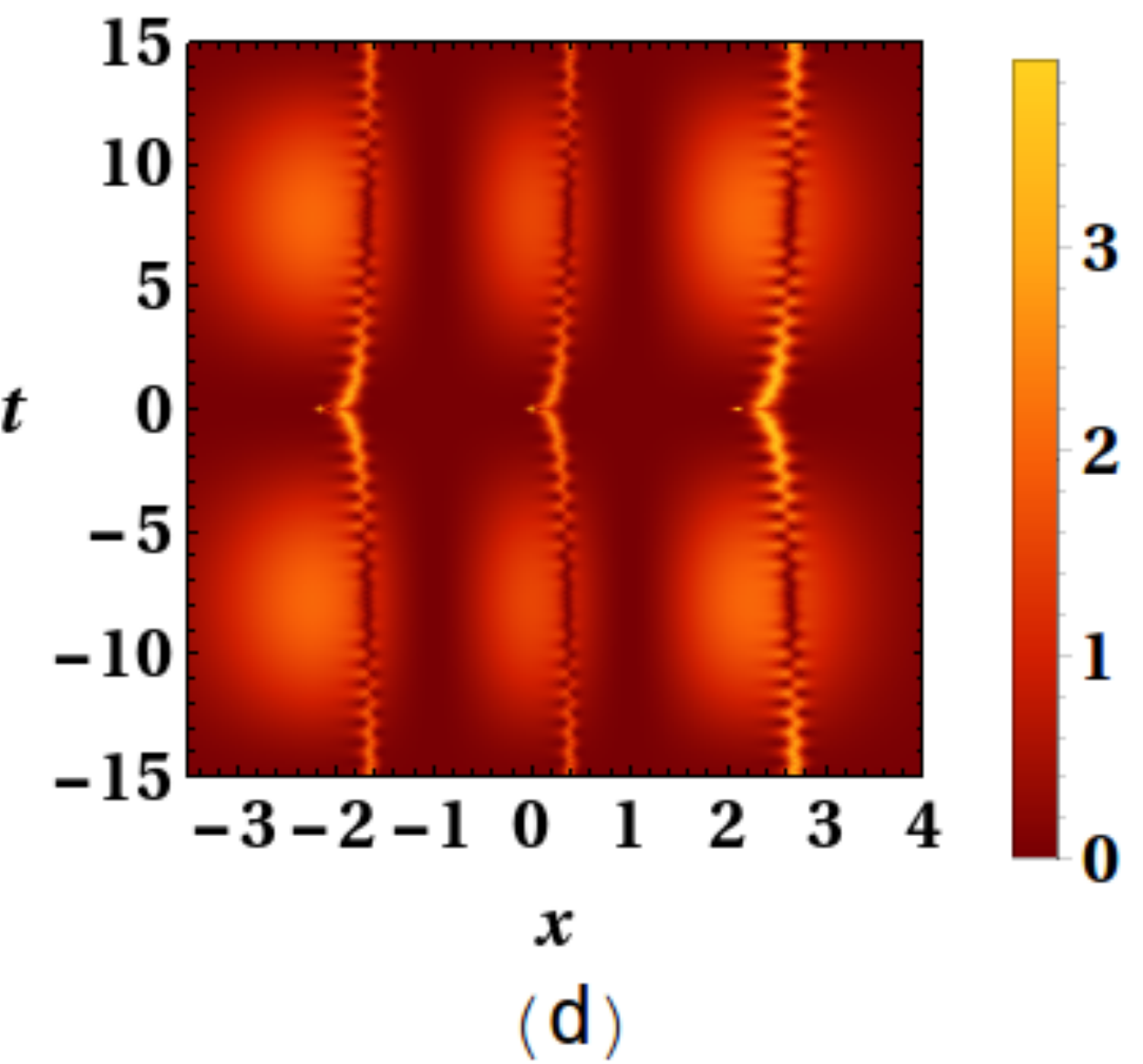}
	\caption{Evolution of two component condensates in spatially modulated 2-CGP system with Rabi coupling for the potential form  given by equation (\ref{pcpo}). The parameters are chosen as $a_{1}=2.8,~a_{2}=0,~f=0.5$,~$n=2$, and $\sigma=0.2$. The left panels show surface plots, while right panels show the  density plots.}
	\label{fig7}
\end{figure}

Here, we construct different types of RW profiles depending on parabolic cylinder potential by varying the quantum model parameter $n$ and also the Rabi term $\sigma$. We would like to recall that the parabolic cylinder function has $(n+1)$ modes as shown in Fig. \ref{fig2}. Such a trapping potential with barrier form can be realized  experimentally \cite{taylor1992,Kevrekidis2007}. In Ref. \cite{zhong2015}, the RW solutions of the two-component Manakov system with variable coefficients have been obtained and an external potential albeit without Rabi coupling. Especially, the pervious study \cite{zhong2015} considered the parametric choice $a_{1}=f\neq0, n=2$ and $a_{2}=0$ without $\sigma$ and shown that the
first component has a bell-shaped peak on top of a long crest extended along the $t$ direction and the second component features a peak with a dip. Here, for comparison purpose we will consider the same parametric choices of $a_{1}, a_{2}, f, n$ at that of Ref.~\cite{zhong2015} but in the presence of Rabi coupling. Interestingly, our study shows that the Rabi coupling induces oscillating boomeronic soliton behaviour. Fig.~\ref{fig7} shows that for $n=2$, we get three number of oscillating boomeronic soliton structures co-existing with bell-shaped peak at the center ($(x,t)=(0,0)$) in the background of oscillating PCM. By increasing $`n'$ we identify that for a given $n$ value there can arise $(n+1)$ oscillating boomeronic solitons. This is the consequence of specific form of the potentials and the production of oscillating boomeronic behaviour is due to the Rabi coupling caused by the applied electro-magnetic field.

\section{Spatially varying three-component Rabi-coupled GP systems}

The three-component generalization of the two coupled GP system (\ref{cgpes}) with Rabi coupling describing the dynamics of three-species condensates in the presence of external electro-magnetic filed can be expressed as

\bea \label{model2}
\fl i\frac{\partial\psi_{j}}{\partial t}+\gamma(x)\frac{\partial^{2}\psi_{j}}{\partial x^{2}}+2g_{jl}(x)\sum^{3}_{l=1}|\psi_{l}|^{2}\psi_{j}+V(x,t)
\psi_{j}-\sum^{3}_{l=1,l\neq j}\sigma\psi_{l}=0,\qquad j=1,2,3,
\eea

where $g_{jl}(x)~(=g(x)),~j,l=1,2,3,$ are spatially varying interacting strengths and $\sigma$ are Rabi coupling term and $V(x,t)$ is the trapping potential. By employing the following unitary transformation in the above equation
\bea \label{unitary3}
\left(\begin{array}{l}
	\psi_{1}(x,t)\\
	\psi_{2}(x,t)\\
	\psi_{3}(x,t)\\
\end{array} \right)=\frac{1}{3}B\left(\begin{array}{l}
	\phi_{1}(x,t)\\
	\phi_{2}(x,t)\\
	\phi_{3}(x,t)\\
\end{array} \right),
\eea
where
\bea B=\left(\begin{array}{ccc}
	(2e^{i\sigma t}+e^{-2i\sigma t})&(e^{-i\sigma t}-e^{i\sigma t})&(e^{-i\sigma t}-e^{i\sigma t})\\
	(e^{-i\sigma t}-e^{i\sigma t})&(2e^{i\sigma t}+e^{-2i\sigma t})&(e^{-i\sigma t}-e^{i\sigma t})\\
	(e^{-i\sigma t}-e^{i\sigma t})&(e^{-i\sigma t}-e^{i\sigma t})&(2e^{i\sigma t}+e^{-2i\sigma t})\\
\end{array} \right),\eea the Rabi coefficient can be absorbed and the resulting inhomogeneous 3-component NLS system can be cast as
\bea\label{cgpe3}
i\frac{\partial\phi_{j}}{\partial t}+\gamma(x)\frac{\partial^{2}\phi_{j}}{\partial x^{2}}+2 g(x) \sum^{3}_{l=1}|\phi_{l}|^{2}\phi_{j}+V(x,t)\phi_{j}=0,~~~~j=1,2,3.
\eea
After performing the similarity transformation given by equation (\ref{trans}) to equation (\ref{cgpe3}), we obtain the following set of
integrable 3-CNLS system
\bea\label{3cnls}
i\frac{\partial q_{j}}{\partial T}+\frac{\partial^{2} q_{j}}{\partial Y^{2}}+2\sum^{3}_{l=1}|q_{l}|^{2}q_{j}=0.~~~~j=1,2,3,
\eea
where $Y$ is given by equation (\ref{YN}) and $T=t$. The soliton solutions of the above system have been obtained in Ref.~\cite{kanna2001} and fascinating
energy sharing collisions have been explored. Then in Ref.~\cite{zhao2013} following explicit three component RW solutions of the above
equation (\ref{3cnls}) were obtained by the DT
\bes\label{3sol}
\bea
q_{1}(Y,T)&=&1-\frac{H_{1}(Y,T)}{G_{1}(Y,T)}~~\exp \left(i \frac{9T}{2}-i \frac{Y}{\sqrt{2}} \right),\\
q_{2}(Y,T)&=&1-\frac{H_{2}(Y,T)}{G_{2}(Y,T)}~~\frac{\exp (5iT)}{\sqrt{2}},\\
q_{3}(Y,T)&=&1-\frac{H_{3}(Y,T)}{G_{3}(Y,T)}~~\exp \left(i \frac{9T}{2}+i \frac{Y}{\sqrt{2}} \right),
\eea \ees
where $H_{j}(Y,T)$ and $G_{j}(Y,T)$, $j=1,2,3,$ are the sixth order polynomial functions given in equation (\ref{3sol}) of Appendix of
Ref. \cite{zhao2013}. It is a straight-forward task to extend our analysis to solitons and other nonlinear wave solutions by this procedure.
This solution has four hidden parameters $A_{1},~A_{2},~A_{3}$ and $A_{4}$, which are also given in Appendix of Ref.~\cite{zhao2013} and due to their
cumbersome form we desist from presenting their explicit forms here again. The different profiles of these RWs and their behavior depend upon
these A-parameters. Particularly, RWs resulting for some acceptable choices of the A-parameters are discussed in Ref.~\cite{zhao2013}.
The $q_{1}$ and $q_{3}$ components of RWs are four-petaled (two humps and two valley) structures while the $q_{2}$ component RW is of eye-shaped
(one hump and two valleys). It is to be noted that by altering A-parameters appropriately one can increase the number of RWs.

Next, we invert the transformations (\ref{trans}) and (\ref{unitary3}) with $j=1,2,3,$ to get the corresponding RW solution of the three-component GP equation
with variable coefficients. This reads as
\bea \label{solu3}
\left(\begin{array}{l}
	\psi_{1}(x,t)\\
	\psi_{2}(x,t)\\
	\psi_{3}(x,t)\\
\end{array} \right)=\frac{N(x)}{3}B\left(\begin{array}{l}
	q_{1}(Y,T)\\
	q_{2}(Y,T)\\
	q_{3}(Y,T)\\
\end{array}\right).
\eea
Here too we consider the same two different forms of the trapping potentials  discussed in previous section.
\subsection{Sech type Hyperbolic Potential (Optical Lattice Potential with $m=1$)}

\begin{figure}[h]
\centering\includegraphics[width=0.95\linewidth]{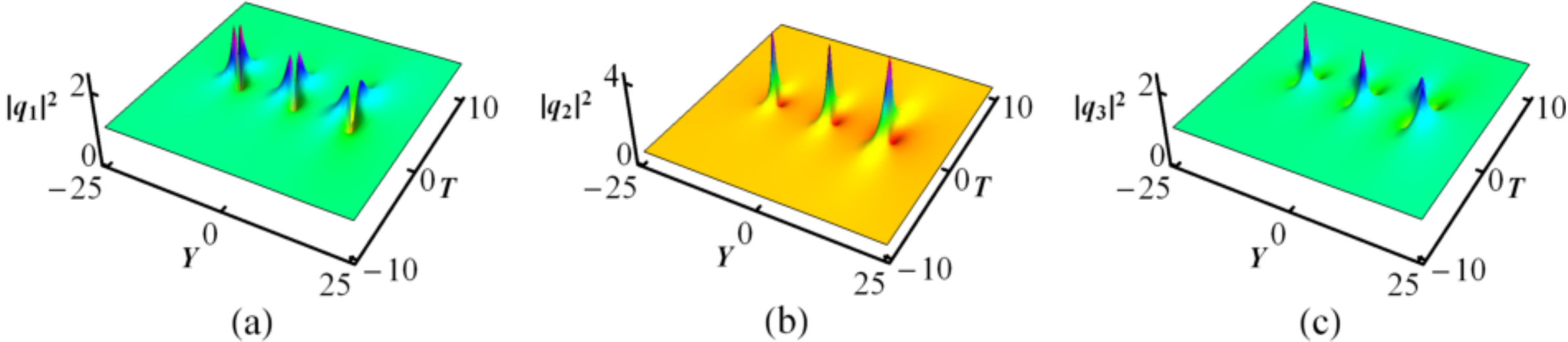}\\
\includegraphics[width=0.95\linewidth]{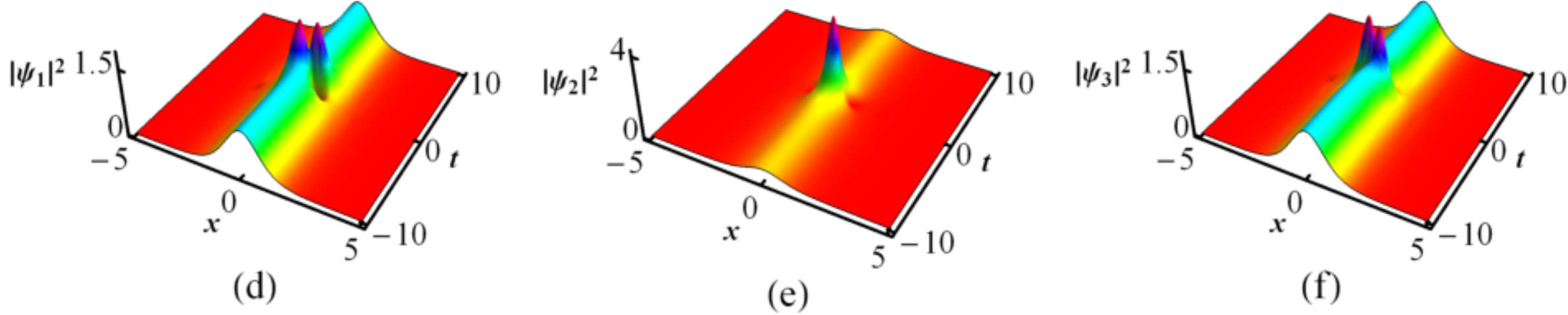}\\
\includegraphics[width=0.95\linewidth]{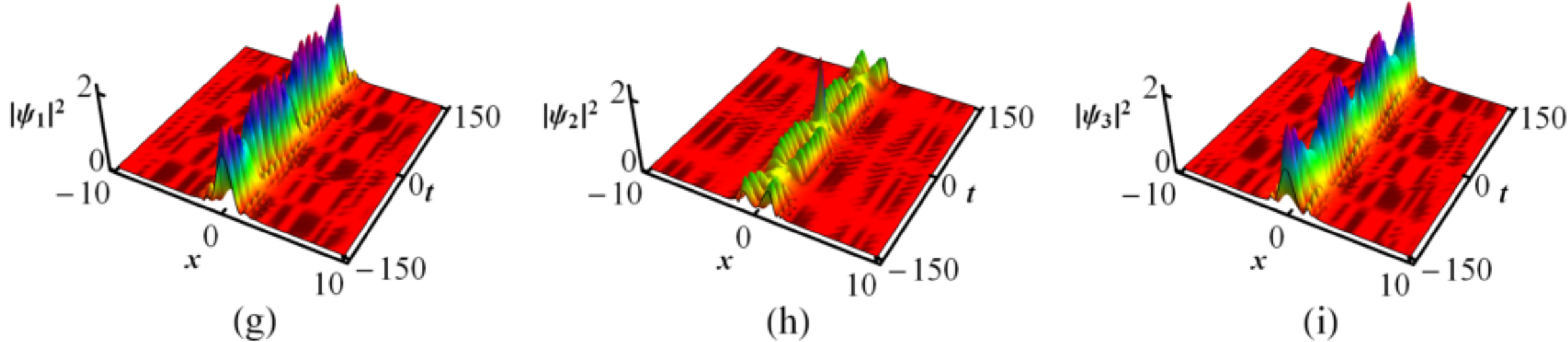}
\caption{Exact rogue wave solutions in system (\ref{model2}) with constant coefficients and in the absence of trapping potential and Rabi coupling
(top panels); spatially modulated coefficients with potential but in the absence of Rabi coupling (middle panels); spatially modulated coefficients with potential and in the presence of Rabi coupling (bottom panels). The parameters
 are chosen as $A_{1}=~A_{3}=0,~A_{2}=25,~A_{4}=-1$, and $\sigma=0.02$.}
\label{fig8}
\end{figure}

\begin{figure}[t]
\centering\includegraphics[width=0.95\linewidth]{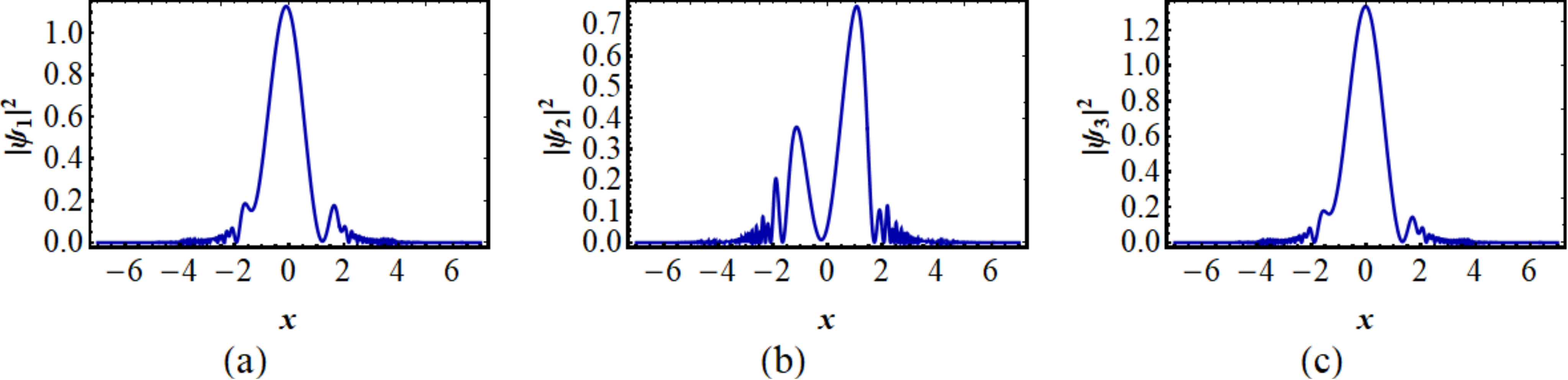}
\caption{Two dimensional plots corresponding to the bottom panels of Fig.~\ref{fig8} for $t=50$.}
\label{fig9}
\end{figure}

First, we consider the trapping potential $V(x)=(1-2 \sech^{2}(x))\sech^{4}(x)$ resulting for $m=1$ in equation (\ref{cnpf})
in the absence of Rabi coupling. For constructing $\psi_{j}$, we assume specific  patterns of RWs in $q_{j}$-s resulting for $A_{1}=A_{3}=0$ and
$A_{2}=A_{4}\neq0~(A_{2}A_{4}<<0)$. For this choice three RWs are localized along time axis as shown in Figs.~\ref{fig8}(a-c) \cite{zhao2013}. The resulting nonlinear
waves with $\sigma=0$ and in the presence of potential and spatial modulation are depicted in Figs.~\ref{fig8}(d-f). In the $\psi_{1}$ component now we have only one zero background RW gets superposed on the bright soliton which broadens the base of the RW and the other two RWs diminish in their amplitudes appear as tiny peaks at wings. The three RWs superposes on a bright soliton around the center (0,0) which broadens the base of the RW and makes the background existing earlier in the absence of potential to vanish. With the introduction of the Rabi coupling (non-zero value for $\sigma$) the amplitude of
the central zero background RW is suppressed considerably and soliton exhibits internal oscillations predominantly at the wings. For example in the $\psi_{1}$ component on can observe periodic recurrence of six solitary waves with oscillating wings. The behaviour of $\psi_{3}$ is similar to that of $\psi_{1}$. These are shown in Figs.~\ref{fig8}(g) and (i) respectively. But in the $\psi_{2}$ component, the internal oscillations of soliton are vibrant resulting in multi-peak solitons as shown in Fig.~\ref{fig8}(h). To facilitate the understanding we separately present the corresponding two-dimensional plots of the three components in Fig.~\ref{fig9}.

\subsection{Parabolic Cylinder Potential in three coupled GP system}

Next, we consider the parabolic cylinder  potential of the form $V(x)= k^{4} (n+\frac{1}{2}-\frac{1}{4}x^{2})D^{4}_{n}(x)$.

\subsubsection*{Case (i)~ $A_{3}=A_{4}=0$:}
\begin{figure}[t]
\centering\includegraphics[width=0.95\linewidth]{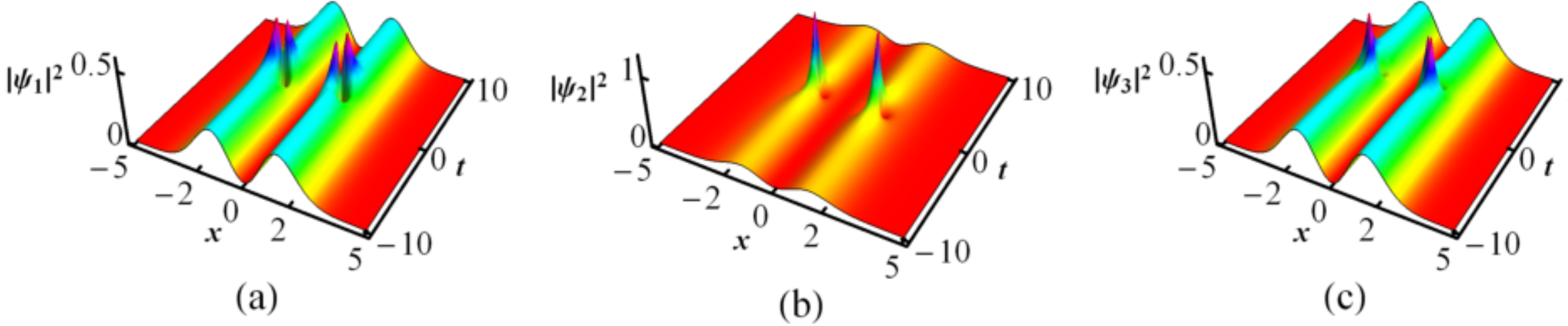}\\
\includegraphics[width=0.95\linewidth]{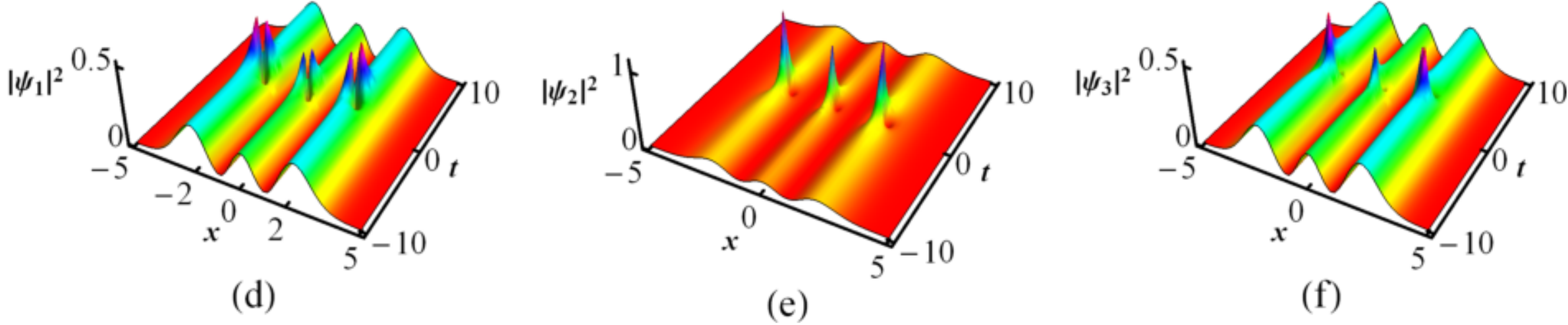}
\caption{Exact rogue wave solutions in the system (\ref{model2}) in the presence of parabolic cylinder potential and in the absence of Rabi coupling
$\sigma=0$. The top panels show the plots for $n=1$ and the bottom panels correspond to $n=2$. The parameter values are $A_{1}=A_{3}=A_{4}=0$ and $ A_{2}=2$.}
\label{fig10}
\end{figure}

In Ref. \cite{zhao2013}, it has been shown that for this choice there arises single RW in the canonical 3-CNLS system (\ref{3cnls}) without Rabi
coupling and trapping potential. In the presence of trapping and inhomogeneities, for the choice $n=1$, a single RW at $t=0$ and its replica appears
at different spatial positions (see at $x=(-2,2)$) as shown in Figs.~\ref{fig10}(a-c). For the same choice, but with $n=2$, as shown in Figs.~\ref{fig10}(d-f) the smaller Parabolic Cylinder Mode (PCM) appears at the center, while the waves with maximum value appear at either sides
of this central PCM. The RWs appear on top of the three PCMs as a result of which we get a shifted three RWs structure.
\subsubsection*{Case (ii)~ $A_{1}\leq0$ and $A_{2}=A_{3}\neq0, A_{4}=0$:}

\begin{figure}[t]
\centering\includegraphics[width=0.95\linewidth]{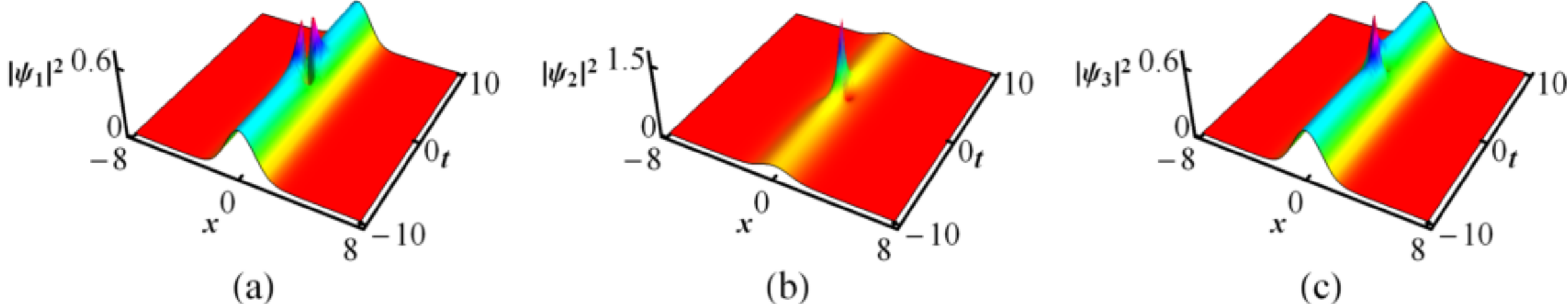}\\
\includegraphics[width=0.95\linewidth]{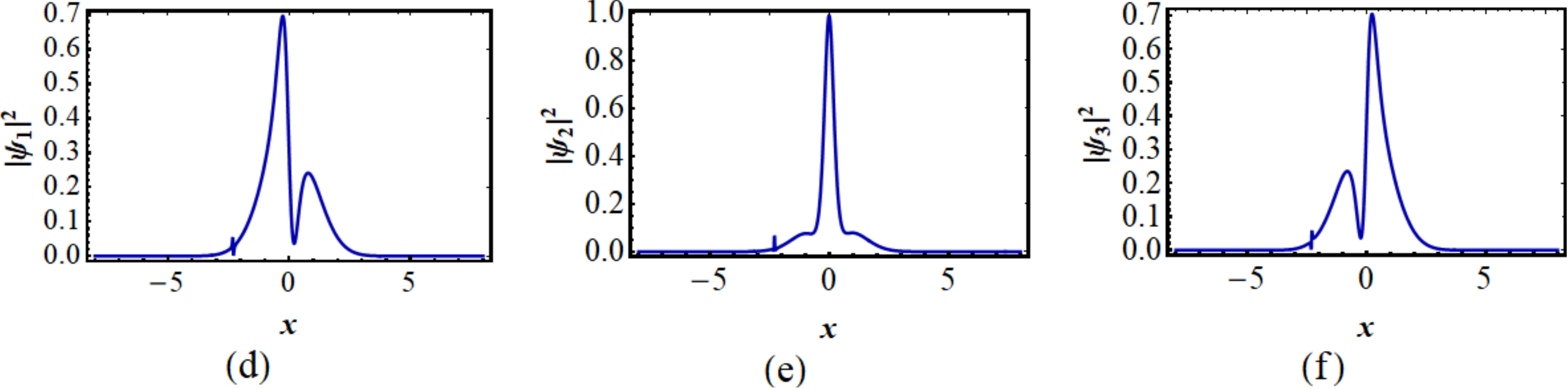}
\caption{Profiles of RWs with parabolic cylindrical potential for $n=0$ and without Rabi coupling. The top panels show the surface plot and the bottom panels show corresponding two dimensional plots for $t=0$ . The system parameters are $A_{1}=A_{4}=0,~A_{2}=30$ and $A_{3}=3$.}
	\label{fig11}
\end{figure}

We consider the parametric choice $A_{1}\leq0$ or $A_{1}\geq0,~A_{2}=A_{3}\neq0,~A_{4}=0$ for which two four petal rogue waves exist in the
3-CNLS system with constant coefficients \cite{zhao2013}. Indeed, here for $A_{1}\leq0$ the two RWs emerge at same time. Now for
the choice of $n=0$, we obtain a zero background rogue wave appearing on top of the zeroth order PCM at $(0,0)$.
The zero background RW appearing in the $-ve$ $x$-axis at $t=0$ has very low intensity at the tail of the PCM. These are shown in the top panels Fig.~\ref{fig11}. The corresponding two dimensional plots are also given in the bottom panels to enhance the understanding of this structure.
\subsubsection*{Case (iii)~ $A_{4}\neq0$:}
\begin{figure}[t]
\centering\includegraphics[width=0.95\linewidth]{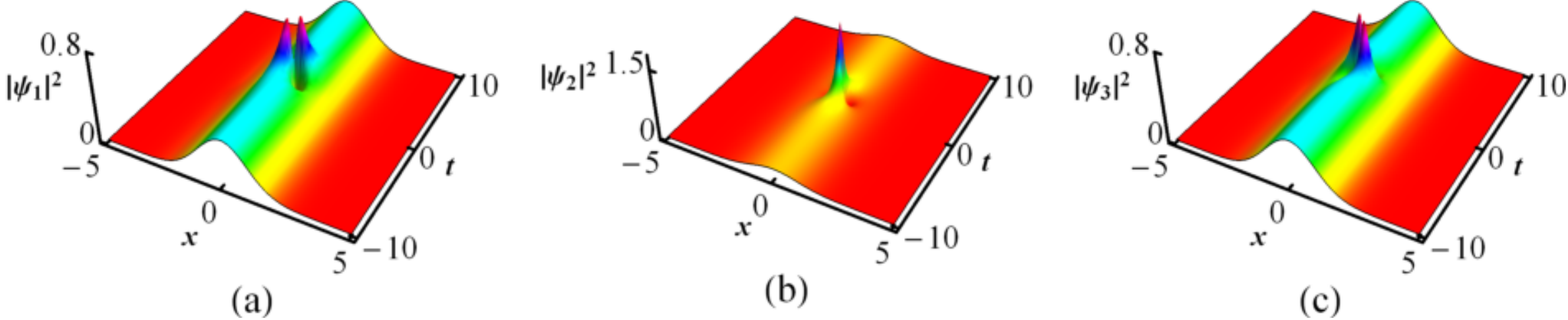}\\
\includegraphics[width=0.95\linewidth]{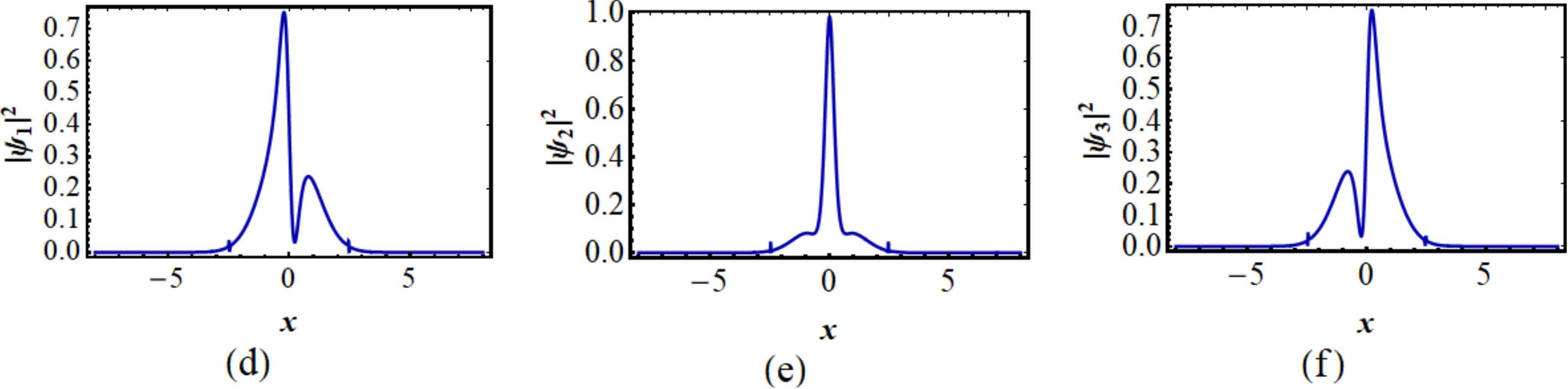}
\caption{ The top panels show the surface plot of exact rogue wave solutions in the system (\ref{model2}) in the presence of parabolic cylinder potential $n=0$ and in the  absence of Rabi coupling( $\sigma=0$). The corresponding two dimensional plots are shown in the bottom panels for $t=0$. The $A$ parameter values are  $A_{1}=0,~A_{2}=120,~A_{3}=0$ and $A_{4}=-1$.}
	\label{fig12}
\end{figure}

\begin{figure}[t]
\centering\includegraphics[width=0.95\linewidth]{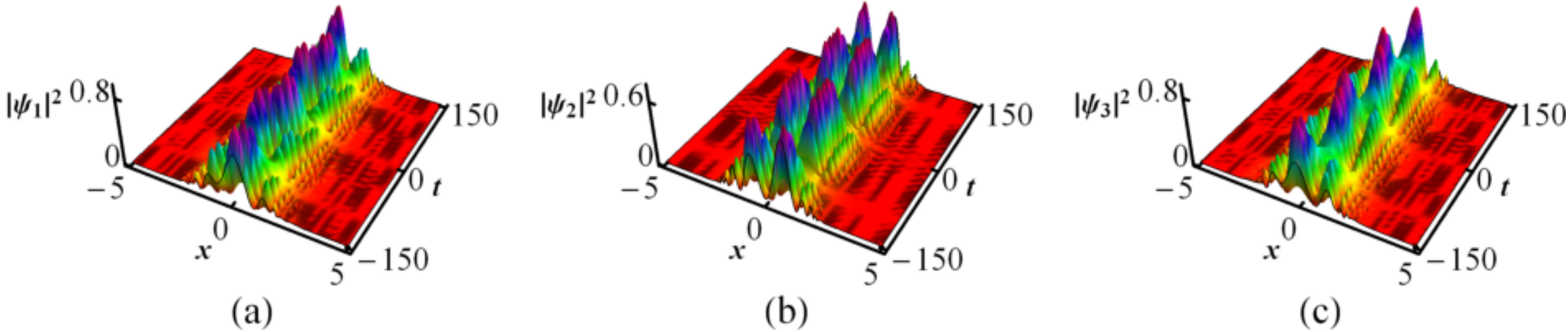}
	\caption{Exact rogue wave solutions in the system (\ref{model2}) in the presence of parabolic cylinder potential $n=0$ and in the  presence of Rabi coupling $\sigma=0.2$ and the $A$ parameter values are $A_{1}=0,~A_{2}=120,~A_{3}=0$ and $A_{4}=-1$.}
	\label{fig13}
\end{figure}

Next, we consider the RW for the choice $A_{4}\neq0$. This choice leads to three rogue wave in constant coefficient 3-CNLS system (see Fig.~\ref{fig12}). For the choice
$n=0$ and  $A_{2} A_{4}<0$, the center of the three RWs displayed in the top panel of Fig.~\ref{fig7} almost coalesces and appears on the top of the PCM as a single shifted RW. As before, here too
for $n=3$ we find an increase in the number of RWs as we move along the spatial direction. Thus from these three cases we note that for arbitrary $n$,
one can have $(n+1)$ RWs appearing on top of the corresponding PCM.
Finally we consider system (\ref{model2}) in the presence of Rabi coupling for any choice of $n$ and $A$ values. Here the internal oscillation with beating behaviour
discussed in Figs.~\ref{fig7}(d-f), is noted predominantly in PCMs as shown Fig.~\ref{fig13}.

\newpage
\section{Conclusion}

In the present work, we have investigated the dynamics of RWs in two- and three- component GP equations with Rabi coupling and spatially varying
parameter functions. The exact solutions have been obtained by using a rotation transformation followed by a similarity transformation. We have
examined the behavior of RWs, for two types of spatially varying trapping potentials which are the lattice potential including the hyperbolic type as special case and parabolic cylinder potential.
To gain insight into the problem first we consider constant coefficient two-component GP system with Rabi coupling and identified the RW co-exists with oscillating
boomeronic solitons. Then we include spatial variation of parameter functions which reveals that new daughter RW with dark (bright) soliton part is produced. Especially, for $\gamma(x)=\sech^{4}(x)$ we demonstrate the possibility of production of oscillating dromions, two-dimensional nonlinear structure in the 2-CGP system. Subsequently, we have studied the interesting behaviour of RWs for spatially varying three-component GP system without Rabi coupling and we have observed different behaviours of RWs. The three RWs are converted into broad based zero background single RW superposed on bright soliton for sech type hyperbolic potential. For parabolic cylinder potential, we observed the appearance of single RW replica in the spatial direction. Then the introduction
of Rabi coupling makes the internal oscillations and beating behaviour in the solitons. Thus our study shows the possibility of formation of 2D structures by spatially varying environment with Rabi coupling even in (1+1) dimensional nonlinear systems. This study can be extended to investigate any kind of nonlinear waves, say solitons, elliptic waves, supported by the
standard integrable CNLS systems and explore their behaviour in the presence of Rabi coupling and spatially varying parameters with different choices of
potentials. Additionally, it is of interest to generalize this study to N-component case with arbitrary $N$ for which rogue wave have been reported recently \cite{rao2019}. This requires the identification of more general unitary transformation. It will be intriguing to study various
modifications of system (\ref{model}) by including spin-orbit coupling, four-wave mixing effects (so-called pair-transition BECs) and also $\cal PT$ symmetric potentials. One can envisage  application of the present study in dispersion management fiber, dispersion control of matter-wave packets and matter wave switches with Rabi coupling. Work is in progress along these directions.

\section*{Acknowledgments}
The work of T. K is supported by Science and Engineering Research Board, Department of Science and
Technology (DST-SERB), Government of India, in the form of a major research project (File No.EMR/2015/
001408). A. A. S acknowledges the financial support from DST in the form of Project Assistant.

\end{document}